\documentstyle[prd,floats,aps]{revtex}

\newcommand{\tg}{\tilde\gamma}
\newcommand{\tG}{\tilde\Gamma}
\newcommand{\tA}{\tilde A}
\newcommand{\tK}{\tilde K}
\newcommand{\lb}{{\cal L}_\beta}
\newcommand{\dt}{(\partial_t - {\cal L}_\beta)\;}

\newcommand{\psibl}{\psi_{BL}}
\newcommand{\bea}{\begin{eqnarray}}
\newcommand{\eea}{\end{eqnarray}}
\newcommand{\beq}{\begin{equation}}
\newcommand{\eeq}{\end{equation}}

\begin{document}

\input{epsf.sty}

\draft

\twocolumn[\hsize\textwidth\columnwidth\hsize\csname
@twocolumnfalse\endcsname

\title{Gauge conditions for long-term numerical black hole evolutions
without excision}

\author{Miguel Alcubierre$^{(1,2)}$, Bernd Br\"ugmann$^{(1)}$, Peter
  Diener$^{(1)}$, Michael Koppitz$^{(1)}$, Denis Pollney$^{(1)}$,
  \\ Edward Seidel$^{(1,3)}$, and Ryoji Takahashi$^{(1,4)}$}

\address{$^{(1)}$ Max-Planck-Institut f\"ur Gravitationsphysik, 
  Albert-Einstein-Institut, Am M\"{u}hlenberg 1, 14476 Golm, Germany \\
  $^{(2)}$ Instituto de Ciencias Nucleares, 
  Universidad Nacional Aut\'onoma de M\'exico, 
  A.P. 70-543, 04510 M\'exico, D. F., M\'exico.
  $^{(3)}$ National Center for Supercomputing Applications,
  Beckman Institute, 405 N. Mathews Ave., Urbana, IL 61801 \\
  $^{(4)}$ Theoretical Astrophysics Center, Juliane Maries Vej 30, 
  2100 Copenhagen, Denmark
  }

\date{June 25, 2002; AEI-2002-037}

\maketitle


\begin{abstract}
  Numerical relativity has faced the problem that standard 3+1
  simulations of black hole spacetimes without singularity excision
  and with singularity avoiding lapse and vanishing shift fail after
  an evolution time of around $30-40M$ due to the so-called slice
  stretching. We discuss lapse and shift conditions for the
  non-excision case that effectively cure slice stretching and allow
  run times of $1000M$ and more.
\end{abstract}


\vskip2pc]


\section{Introduction}

A crucial role in numerical relativity simulations of two black holes
(BH) is played by the choice of coordinates. This gauge choice
involves both the choice of a lapse function and of a shift vector,
which typically have to be determined dynamically during a numerical
evolution.  The first results for colliding BH's were obtained for
head-on collisions using the ADM decomposition of the Einstein
equations with the lapse determined by the maximal slicing condition
and the shift vector set to zero~\cite{Smarr75,Smarr77,Anninos93b}.
Maximal slices are known to be singularity avoiding, that is, starting
from BH initial data where the physical singularity is to the future
of the initial hypersurface, the lapse approaches the Minkowski value
of unity in the asymptotically flat regions, but approaches zero near
the physical singularity.  In this way one can in principle foliate a
BH spacetime without singularities, but since time marches on in the
far regions while being frozen in the interior, the slices become more
and more distorted. Historically, this phenomenon has been called
`grid stretching' by the numerical relativity community, though we
will refer to it as `slice stretching' since it is a property of the
slices themselves, quite independent of the existence of a numerical
grid.  Slice stretching introduces a difficult problem for numerical
simulations since the metric develops large gradients that keep on
growing until the numerical code can no longer handle them and fails.
Advanced numerical methods can help in spherical symmetry, see,
e.g.,~\cite{Bona94b}, but to date have not proved very successful in
three-dimensional (3D) evolutions~\cite{Walker98a}.

Nonetheless, such singularity avoiding slicings with vanishing shift
have been quite successful, since they do allow a finite evolution
time of roughly $t=30-40M$ (with $M$ the total ADM mass of the system)
for fully 3D evolutions of BH's, as first demonstrated in 1995 for the
case of a single Schwarzschild BH~\cite{Anninos94c}.
In~\cite{Bruegmann97} the first fully 3D simulation of the grazing
collision of two nearby BH's was performed with singularity avoiding
slicing and vanishing shift, lasting for about $7M$. With improved
techniques the grazing collision has recently been pushed to about
$35M$, which for the first time allowed the extraction of
gravitational waveforms from a 3D numerical
merger~\cite{Alcubierre00b}.  And even though singularity avoiding
slicings with vanishing shift have so far been limited to a finite
time interval of only $30-40M$, this interval can be moved into the
truly non-linear regime of a plunge starting from an approximate
innermost stable circular orbit (ISCO) of two BH's, since the
remainder of the merger and ring-down can be computed using the close
limit approximation~\cite{Baker00b}.  Following such an approach, the
first waveforms for the plunge from an approximate ISCO have been
obtained~\cite{Baker:2001nu,Baker:2001sf,Baker:2002qf}.

So far the most important strategy to avoid slice stretching has been
black hole excision~\cite{Thornburg87,Seidel92a}.  The idea is to use
horizon penetrating coordinates (notice that maximal slicing {\em is}
horizon penetrating unless one imposes the extra boundary condition of
having a vanishing lapse at the horizon) and to excise the interior of
the BH's from the numerical grid. A non-vanishing shift is essential
to keep grid points from falling into the BH.  This approach has seen
many successful implementations for single black holes.  First
demonstrated in 3D in~\cite{Anninos94c}, with further development it
has, in particular, allowed to move a black hole across the numerical
grid~\cite{Gomez98a}.  If a stable numerical implementation can be
found, this approach should make it possible to simulate many orbits
of two well separated BH's.  The key difference between BH excision
and the use of singularity avoiding slicings with vanishing shift is
that with excision single static BH's can be stably evolved for
thousands of $M$; see~\cite{Gomez98a} for the case of evolutions using
null coordinates (that do not directly generalize to binary BH
systems), and recently $100,000M$ have been reached for a single BH
with a 3+1 Cauchy code in octant symmetry~\cite{Alcubierre00a}.  Black
hole excision holds a lot of promise, even if currently evolutions of
only $9$--$15M$ have been achieved for binary BH's~\cite{Brandt00}.

In this paper we demonstrate that the new lapse and shift conditions
introduced in~\cite{Alcubierre01a} for the case of a single distorted
BH with excision (using the excision techniques
of~\cite{Alcubierre00a}) can work well even {\em without
excision}. This allows us to break through the barrier of about $35M$
for singularity avoiding slicings in 3+1 numerical relativity.  Our
gauge choice maintains singularity avoidance but cures the main
problems associated with slice stretching, allowing us to reach $500M$
and more for the evolution of single or even distorted BH's. For BH's
colliding head-on that merge early on during the evolution, i.e.\
which start out sufficiently close to each other, the final BH can
again be evolved for hundreds or even thousands of $M$.

Moreover, these gauge conditions have two important effects: (a) they
drive the system towards a static state, virtually, if not completely,
eliminating the chronic growth in metric functions typical of slice
stretching.  Hence, in principle they should allow for indefinitely
long evolutions (if no other instabilities develop; see below).  (b)
since unbounded growth in metric functions is halted, they allow {\em
much more accurate} results to be obtained for extremely long times,
and at lower resolution than before.  Below we will show results
obtained for colliding black holes that show only 10\% error in the
horizon mass after more than $5000M$ of evolution.

The evolutions in this paper are carried out using the puncture method
for evolutions~\cite{Bruegmann97,Anninos94c}.  The starting point is
initial data of the Brill-Lindquist topology in isotropic
coordinates~\cite{Brill63}.  This `puncture' data is defined on $R^3$
minus a point for each of the internal asymptotically flat ends of the
BH's.  If one treats the coordinate singularity at the punctures
appropriately, the punctures do not evolve as long as the shift
vanishes there.  That is, the metric and the extrinsic curvature do
not evolve at the punctures.  It can also be checked that the
maximal slicing equation produces a smooth numerical solution for the
lapse at the punctures.

One basic observation for our choice of shift vector is that the ``Gamma
freezing'' shift introduced in~\cite{Alcubierre01a} for our project in
simple BH excision has the following property when the BH's are not
excised but are represented as punctures:  Initially the shift is
zero, but as the slice stretching develops, the shift reacts by
pulling out points from the inner asymptotically flat region near the
punctures.  The lapse and shift conditions taken together are then
able to virtually stop the evolution of one or even two black holes,
essentially mimicking the behavior of the lapse and shift known from
stable evolutions of a BH in Kerr-Schild coordinates.  This is a key 
result that will be detailed below.

The paper is organized as follows.  First we introduce the evolution
equations and the constraints in Section~\ref{se:formulation}. 
In Sec.~\ref{se:gauge} we discuss the gauge conditions. 
The puncture initial data and puncture evolutions are discussed in 
Sec.~\ref{se:punctures} and Sec.~\ref{se:gaugepuncevol}.
In Sec.~\ref{se:numerics} miscellaneous aspects of our numerical
implementation are discussed.
In Sec.~\ref{se:applications} we present numerical results for one and
two BH's, and we conclude in Sec.~\ref{se:conclusion}.


\section{Formulation}
\label{se:formulation}

The standard variables in the 3+1 formulation of ADM
(Arnowitt-Deser-Misner, see~\cite{York79}) are the 3-metric
$\gamma_{ij}$ and the extrinsic curvature $K_{ij}$. The gauge is
determined by the lapse function $\alpha$ and the shift vector
$\beta^i$. We will only consider the vacuum case. The evolution
equations are
\bea
\dt \gamma_{ij} &=& - 2 \alpha K_{ij},
\label{dgdt} \\
\dt K_{ij} &=& - D_i D_j \alpha 
\nonumber \\
&&  + \alpha (R_{ij} + K K_{ij} - 2 K_{ik} {K^k}{}_j),
\label{dKdt}
\eea
and the constraints are
\bea
{\cal H} & \equiv & R + K^2 - K_{ij} K^{ij} = 0,
\label{Hconstraint}\\
{\cal D}^i & \equiv & D_j (K^{ij} - \gamma^{ij} K) = 0.
\label{Dconstraint}
\eea Here $\lb$ is the Lie derivative with respect to the shift vector
$\beta^i$, $D_i$ is the covariant derivative associated with the
3-metric $\gamma_{ij}$, $R_{ij}$ is the three-dimensional Ricci
tensor, $R$ the Ricci scalar, and $K$ is the trace of $K_{ij}$.

We will use the BSSN form of these equations (Baumgarte, Shapiro
\cite{Baumgarte99}, and Shibata, Nakamura \cite{Shibata95}). One
introduces new variables based on a trace decomposition of the
extrinsic curvature and a conformal rescaling of both the metric and
the extrinsic curvature. The trace-free part $A_{ij}$ of the extrinsic
curvature is defined by
\beq
A_{ij} = K_{ij} - \frac{1}{3} \gamma_{ij} K. 
\label{defA}
\eeq
Assuming that the metric $\gamma_{ij}$ is obtained from a conformal
metric $\tg_{ij}$ by a conformal transformation,
\beq 
\gamma_{ij} = \psi^4 \tg_{ij}, 
\eeq 
we can choose a conformal
factor $\psi$ such that the determinant of $\tg_{ij}$ is 1: 
\bea
\psi & = & \gamma^{1/12},
\\
\tg_{ij} & = & \psi^{-4}\gamma_{ij} = \gamma^{-{1/3}}\gamma_{ij},
\\
\tilde\gamma &=& 1,
\eea
where $\gamma$ is the determinant of $\gamma_{ij}$ and $\tilde\gamma$
is the determinant of $\tg_{ij}$. Instead of $\gamma_{ij}$ and $K_{ij}$
we can therefore use the variables
\bea
  \phi &=& \ln \psi = \frac{1}{12} \ln \gamma,
\\
  K &=& \gamma_{ij} K^{ij},
\\
  \tg_{ij} &=& e^{-4\phi} \gamma_{ij},
\\
  \tA_{ij} &=& e^{-4\phi} A_{ij},
\eea
where $\tg_{ij}$ has determinant 1 and $\tA_{ij}$ has vanishing trace.
Furthermore, we introduce the conformal connection functions
\beq
  \tG^i = \tg^{jk} \tG^i{}_{jk} = - \partial_j \tg^{ij},
\label{defG}
\eeq where $\tG^i{}_{jk}$ is the Christoffel symbol of the conformal
metric and where the second equality holds only if the determinant of
the conformal 3-metric $\tilde \gamma$ is unity (which is true
analytically but may not hold numerically). We call $\phi$, $K$,
$\tg_{ij}$, $\tA_{ij}$, and $\tG^i$ the BSSN variables.

In terms of the BSSN variables the evolution equation (\ref{dgdt})
becomes
\bea
  \dt \tg_{ij} &=& -2 \alpha \tA_{ij},  
\label{dtgdt}
\\
  \dt \phi &=& - \frac{1}{6} \alpha K,
\label{dphidt}
\eea
while equation (\ref{dKdt}) leads to
\bea
  \dt \tA_{ij} &=& e^{-4\phi} [ - D_i D_j \alpha + \alpha R_{ij}]^{TF}
\nonumber \\
  && + \;\alpha (K \tA_{ij} - 2 \tA_{ik} \tA^k{}_j),
\label{dtAdt}
\\
  \dt K &=& - D^i D_i \alpha
\nonumber \\
  && + \;\alpha (\tA_{ij} \tA^{ij} + \frac{1}{3} K^2),  
\label{dtrKdt}
\eea
where $TF$ denotes the trace-free part of the expression in brackets
with respect to $\gamma_{ij}$. Note that the right-hand side of the
evolution equation (\ref{dtAdt}) for the trace-free variable $\tA_{ij}$
is trace-free except for the term $\tA_{ik} \tA^k{}_j$. This
is no contradiction since the condition that $\tA_{ij}$ remains trace-free 
is 
$\dt (\tg^{ij}\tA_{ij}) = 0$ and not $\tg^{ij}\dt \tA_{ij} = 0$.

On the right-hand side of (\ref{dtrKdt}) we have used the Hamiltonian
constraint (\ref{Hconstraint}) to eliminate the Ricci scalar,
\beq
  R = K_{ij} K^{ij} - K^2 = \tA_{ij} \tA^{ij} - \frac{2}{3} K^2.
\eeq
The momentum constraint~(\ref{Dconstraint}) becomes
\beq
  \partial_j \tA^{ij} = - \tG^i{}_{jk} \tA^{jk} - 6 \tA^{ij}
                        \partial_j \phi + \frac{2}{3} \tg^{ij}
                        \partial_j K.
\eeq

An evolution equation for $\tG^i$ can be obtained from 
(\ref{defG}) and (\ref{dtgdt}),
\bea
\partial_t \tG^i 
&=&
- 2 \left( \alpha \partial_j \tA^{ij} + \tA^{ij} \partial_j \alpha \right)
- \partial_j \left( \lb \tg^{ij} \right) ,
\label{dtGdt0}
\eea
where we will use the momentum constraint above to substitute for
the divergence of $\tA^{ij}$. One subtlety in obtaining numerically
stable evolutions with the BSSN variables is precisely the question of
how the constraints are used in the evolution equations. Several
choices are possible and have been studied, see
e.g.~\cite{Alcubierre99e}.

Note that in the preceding equations we are computing Lie derivatives
of tensor densities. If the weight of a tensor density $T$ is $w$,
i.e.\ if $T$ is a tensor times $\gamma^{w/2}$, then
\beq \lb T = [\lb
T]^{w=0}_\partial + w T \partial_k\beta^k,
\eeq
where the first term denotes the tensor formula for Lie derivatives
with the derivative operator $\partial$ and the second is the
additional contribution due to the density factor.  The density weight
of $\psi = e^\phi$ is $\frac{1}{6}$, so the weight of $\tg_{ij}$ and
$\tA_{ij}$ is $-\frac{2}{3}$ and the weight of $\tg^{ij}$ is
$\frac{2}{3}$. To be explicit, \bea \lb \phi &=& \beta^k\partial_k\phi
+ \frac{1}{6}\partial_k\beta^k,
\label{lbphi}
\\
\lb \tg_{ij} &=& \beta^k\partial_k\tg_{ij} + 
                 \tg_{ik}\partial_j\beta^k +
                 \tg_{jk}\partial_i\beta^k - 
                 \frac{2}{3} \tg_{ij}\partial_k\beta^k,
\\
\lb \tg^{ij} &=& \beta^k\partial_k\tg^{ij} - 
                 \tg^{ik}\partial_k\beta^j -
                 \tg^{jk}\partial_k\beta^i + 
                 \frac{2}{3} \tg^{ij}\partial_k\beta^k.
\eea
The evolution equation (\ref{dtGdt0}) for $\tG^i$ therefore becomes
\bea
\partial_t \tG^i &=&
\tilde\gamma^{jk} \partial_j\partial_k \beta^i
+ \frac{1}{3} \tilde\gamma^{ij}  \partial_j\partial_k\beta^k
\nonumber \\
&& 
+ \beta^j\partial_j \tilde\Gamma^i
- \tilde\Gamma^j \partial_j \beta^i 
+ \frac{2}{3} \tilde\Gamma^i \partial_j\beta^j
\nonumber \\
&& - 2 \tilde{A}^{ij} \partial_j\alpha
+ 2 \alpha ( 
\tilde{\Gamma}^i{}_{jk} \tilde{A}^{jk} + 6 \tilde{A}^{ij}
\partial_j \phi - \frac{2}{3} \tg^{ij} \partial_j K
).
\nonumber \\
\label{dtGdt}
\eea In the second line we see the formula for a vector density of
weight $\frac{2}{3}$, but since $\tG^i$ is not really a tensor density
but is derived from Christoffel symbols we obtain extra terms
involving second derivatives of the shift (the first line in the
equation above).

On the right-hand sides of the evolution equations for $\tA_{ij}$ and
$K$, (\ref{dtAdt}) and (\ref{dtrKdt}), there occur covariant
derivatives of the lapse function, and the Ricci tensor of the
non-conformal metric.  Since
\beq
\Gamma^k{}_{ij} = \tG^k{}_{ij} + 2 (
\delta^k_i \partial_j\phi + \delta^k_j\partial_i\phi - \tg_{ij}
\tg^{kl} \partial_l\phi),
\label{Christoffel}
\eeq
where $\tG^k{}_{ij}$ is the Christoffel symbol of the conformal
metric, we have for example
\beq
D^i D_i \alpha = e^{-4\phi} (\tg^{ij} \partial_i\partial_j\alpha -
\tG^k\partial_k\alpha + 2\tg^{ij} \partial_i\phi \partial_j\alpha).
\eeq
The Ricci tensor can be separated in two parts,
\beq
R_{ij} = \tilde{R}_{ij} + R^{\phi}_{ij},
\label{Ricci}
\eeq
where $\tilde{R}_{ij}$ is the Ricci tensor of the
conformal metric and $R^{\phi}_{ij}$ denotes additional terms
depending on $\phi$:
\bea
R^{\phi}_{ij} &=& - 2 \tilde{D}_i
\tilde{D}_j \phi - 2 \tilde{\gamma}_{ij}
\tilde{D}^k \tilde{D}_k \phi \nonumber \\
&& + 4 \tilde{D}_i \phi \; \tilde{D}_j \phi - 4 \tilde{\gamma}_{ij}
\tilde{D}^k \phi \; \tilde{D}_k \phi ,
\label{Ricciphi}
\eea
with $\tilde D_i$ the covariant derivative associated with the
conformal metric.  The conformal Ricci tensor can be written in terms
of the conformal connection functions as
\bea
\tilde R_{ij} & = & - \frac{1}{2} \tg^{lm} \partial_l \partial_m
\tg_{ij} + \tg_{k(i} \partial_{j)} \tG^k
+ \tG^k \tG_{(ij)k} \nonumber \\
& & + \tg^{lm} \left( 2 \tG^k{}_{l(i} \tG_{j)km} + \tG^k{}_{im}
  \tG_{klj} \right).
\label{Ricciconf}
\eea

A key observation here is that if one introduces the $\tG^i$ as
independent variables, then the principal part of the right-hand side
of equation (\ref{dtAdt}) contains the wave operator $\tg^{lm}
\partial_l \partial_m \tg_{ij}$ but no other second derivatives of the
conformal metric. This brings the evolution system one step closer to
being hyperbolic. 

One of the reasons why we have written out the BSSN system in such
detail is to point out a subtlety that arises in the actual
implementation if one wants to achieve numerical stability. In the
computer code we do not use the numerically evolved $\tG^i$ in all
places, but follow this rule:
\begin{itemize}
\item Partial derivatives $\partial_j\tG^i$ are computed as finite
  differences of the independent variables $\tG^i$ that are evolved
  using (\ref{dtGdt}).
\item In all expressions that just require $\tG^i$ and not its
derivative we substitute $\tg^{jk}\tG^i{}_{jk}(\tg)$, that is we do
not use the independently evolved variable $\tG^i$ but recompute
$\tG^i$ according to its definition (\ref{defG}) from the current
values of $\tg_{ij}$.  
\end{itemize}
In practice we have found that the evolutions are far less stable if
either $\tG^i$ is treated as an independent variable everywhere, or if
$\tG^i$ is recomputed from $\tg_{ij}$ before each time step. The rule
just stated helps to maintain the constraint $\tG^i = -
\partial_j\tg^{ij}$ well behaved without removing the advantage of
reformulating the principal part of the Ricci tensor.

In summary, we evolve the BSSN variables $\tg_{ij}$, $\phi$,
$\tA_{ij}$, $K$, and $\tG^i$ according to (\ref{dtgdt}),
(\ref{dphidt}), (\ref{dtAdt}), (\ref{dtrKdt}), and (\ref{dtGdt}),
respectively. The Ricci tensor is separated as shown in (\ref{Ricci})
with each part computed according to (\ref{Ricciphi}) and
(\ref{Ricciconf}) respectively. The Hamiltonian and momentum
constraints have been used to write the equations in a particular way.
The evolved variables $\tG^i$ are only used when their partial
derivatives are needed (the one term in the conformal Ricci tensor
(\ref{Ricciconf}) and the advection term $\beta^k \partial_k \tG^i$ in
the evolution equation for the $\tG^i$ themselves, Eq.~(\ref{dtGdt})).


\section{The Gauge Conditions}
\label{se:gauge}

We will consider families of gauge conditions that are not restricted
to puncture data and that can be used in
principle with any 3+1 form of the Einstein's equations that allows a
general gauge.  However, the specific family we test in this paper is
best motivated by considering the BSSN system introduced above.  For
the present purposes, of special importance are the following two
properties of this formulation:

\begin{itemize}
\item The trace of the extrinsic curvature $K$ is treated as an
  independent variable.  For a long time it has been known that the
  evolution of $K$ is directly related to the choice of a lapse
  function $\alpha$.  Thus, having $K$ as an independent field allows
  one to impose slicing conditions in a much cleaner way.
\item The appearance of the ``conformal connection functions''
  $\tilde\Gamma^i$ as independent quantities.  As already noted by
  Baumgarte and Shapiro~\cite{Baumgarte99} (see
  also~\cite{Andersson98,Andersson01a}), the evolution equation for these
  quantities can be turned into an elliptic condition on the shift
  which is related to the minimal distortion condition.  More
  generally, one can relate the shift choice to the evolution of these
  quantities, again allowing for a clean treatment of the shift
  condition.
\end{itemize}

Our aim is to look for gauge conditions that at late times, once the
physical system under consideration has settled to a final stationary
state, will be able to drive the coordinate system to a frame where
this stationarity is evident.  In effect, we are looking for
``symmetry seeking'' coordinates of the type discussed by Gundlach and
Garfinkle and also by Brady, Creighton, and
Thorne~\cite{Garfinkle99,Brady_P:98} that will be able to find the
approximate Killing field that the system has at late times.  In order
to achieve this we believe that the natural approach is to relate the
gauge choice to the evolution of certain combinations of dynamic
quantities in such a way that the gauge will either freeze completely
the evolution of those quantities (typically by solving some elliptic
equations), or will attempt to do so with some time delay (by solving
instead parabolic or hyperbolic equations).

We will consider the lapse and shift conditions in turn.  Special
cases of the gauge conditions that we will introduce here were
recently used together with BH excision with remarkable results
in~\cite{Alcubierre00a}, but as we will show below, the conditions are
so powerful that in the cases tested, they work {\em even without excision}.


\subsection{Slicing conditions}
\label{sec:lapse}

The starting point for our slicing conditions is the ``$K$-freezing''
condition $\partial_t K$=0, which in the particular case when $K$=0
reduces to the well known ``maximal slicing'' condition.  The
$K$-freezing condition leads to the following elliptic equation for the
lapse:
\begin{equation}
\Delta \alpha = \beta^i \partial_i K + \alpha K_{ij} K^{ij} \, ,
\label{eq:Kfreezing}
\end{equation}
with $\Delta$ the Laplacian operator for the spatial metric
$\gamma_{ij}$.  In the BSSN formulation, once we have solved the
elliptic equation for the lapse, the $K$-freezing condition can be
imposed at the analytic level by simply not evolving $K$.

One can construct parabolic or hyperbolic slicing conditions by
making either $\partial_t \alpha$ or $\partial^2_t \alpha$ proportional to
$\partial_t K$.  We call such conditions ``K-driver'' conditions
(see~\cite{Balakrishna96a}).  The hyperbolic K-driver condition has
the form~\cite{Bona94b,Alcubierre01a}
\begin{eqnarray}
\partial_t \alpha &=& - \alpha^2 f(\alpha) (K - K_0), 
\label{eq:hypKdriver}
\end{eqnarray}
where $f(\alpha)$ is an arbitrary positive function of $\alpha$ and
$K_0 = K(t=0)$.  In
our evolutions, we normally take 
\beq
f(\alpha)=\frac{2}{\alpha}, 
\eeq
which is referred to as 1+log slicing, since empirically we have found
that such a choice has excellent singularity avoiding properties. In
Sec.~\ref{se:puncevol} we introduce a modification of $f(\alpha)$ for
puncture evolutions.  The hyperbolic K-driver condition is in fact
only a slight generalization of the Bona-Masso family of slicing
conditions~\cite{Bona94b}: $\partial_t \alpha = - \alpha^2 f(\alpha)
K$.

By taking an extra time derivative of the slicing condition above, and
using the evolution equation for $K$, one can see that the lapse obeys
a generalized wave equation,
\bea
\partial^2_t \alpha &=& - \partial_t (\alpha^2 f) (K - K_0) - 
	\alpha^2 f \partial_t K \nonumber
\\
&=& \alpha^2 f (\Delta \alpha - \alpha K_{ij} K^{ij} - \beta^iD_iK
	+ 2\alpha f + \alpha^2 f').
\label{wavelapse}
\eea
Previously we have also experimented with a somewhat different
form of hyperbolic K-driver condition
\beq \partial^2_t \alpha = -
\alpha^2 f \partial_t K,
\label{eq:hypKdriver2}
\eeq where the right hand side vanishes in the case that K freezing is
achieved. However, one may anticipate the problem that even in the case when
$\partial_t K = 0$ we only obtain $\partial_t\alpha = const$, while
for (\ref{eq:hypKdriver}) we see that $K = K_0$ implies that
$\partial_t\alpha = 0$.  Moreover, in black hole evolutions where the
lapse collapses to zero, condition~(\ref{eq:hypKdriver}) guarantees
that the lapse will stop evolving, while
condition~(\ref{eq:hypKdriver2}) only implies that $\partial_t\alpha$
will stop evolving so the lapse can easily ``collapse'' beyond zero
and become negative.  For these reasons, in practice the condition
Eq.~(\ref{eq:hypKdriver}) leads to more stable black hole evolutions,
and this is the slicing condition that we consider in this paper.

The wave speed in both cases is $v_\alpha = \alpha \sqrt{f(\alpha)}$,
which explains the need for $f(\alpha)$ to be positive. Notice that,
depending on the value of $f(\alpha)$, this wave speed can be larger
or smaller than the physical speed of light.  This represents no
problem, as it only indicates the speed of propagation of the
coordinate system, i.e.\ it is only a ``gauge speed''.  In particular,
for the 1+log slicing introduced above with $f=2/\alpha$, the gauge
speed in the asymptotic regions (where $\alpha \simeq 1$) becomes
$v_\alpha=\sqrt{2}>1$.  One could then argue that choosing
$f=1/\alpha$ should be a better alternative, as the asymptotic gauge
speed would then be equal to the physical speed of light.  However,
experience has shown that such a choice is not nearly as robust and
seems to lead easily to gauge pathologies as those studied
in~\cite{Alcubierre97a,Alcubierre97b}.


\subsection{Shift conditions}
\label{sec:shift}

In the BSSN formulation, an elliptic shift condition is easily
obtained by imposing the ``Gamma-freezing'' condition $\partial_t
\tilde\Gamma^k$=0, or using (\ref{dtGdt}),
\begin{eqnarray}
  \tilde\gamma^{jk} \partial_j\partial_k \beta^i
+ \frac{1}{3} \tilde\gamma^{ij}  \partial_j\partial_k\beta^k
- \tilde\Gamma^j \partial_j \beta^i 
+ \frac{2}{3} \tilde\Gamma^i \partial_j\beta^j
+ \beta^j\partial_j \tilde\Gamma^i
&& \nonumber \\
- 2 \tilde A^{ij} \partial_j\alpha
- 2 \alpha \left( \frac{2}{3}\tilde\gamma^{ij}\partial_j K 
- 6 \tilde A^{ij}\partial_j\phi - \tilde \Gamma^i_{jk} \tilde A^{jk} \right)
 = 0 .
&& \nonumber \\
&& \hspace*{-3cm}
\label{eq:Gammafreezing}
\end{eqnarray}
Notice that, just as it happened with the $K$-freezing condition for
the lapse, once we have solved the previous elliptic equations for the
shift, the Gamma-freezing condition can be enforced at an analytic
level by simply not evolving the $\tilde\Gamma^k$.

The Gamma-freezing condition is closely related to the well known
minimal distortion shift condition~\cite{Smarr78b}.  In order to see
exactly how these two shift conditions are related, we write here the
minimal distortion condition
\begin{equation}
\nabla_j \Sigma^{ij} = 0 \, , 
\label{eq:mindis}
\end{equation}
where $\Sigma_{ij}$ is the so-called ``distortion tensor'' defined as
\begin{equation}
\Sigma_{ij} := \frac{1}{2} \gamma^{1/3} \partial_t \tilde{\gamma}_{ij} \, ,
\label{eq:distortion}
\end{equation}
with $\tilde{\gamma}_{ij}$ the same as before.  A little algebra
shows that the evolution equation for the conformal connection
functions~(\ref{dtGdt}) can be written in terms of $\Sigma_{ij}$ as
\begin{equation}
\partial_t \tilde{\Gamma}^i = 2 \partial_j \left( \gamma^{1/3} \Sigma^{ij}
\right) \, .
\end{equation}
More explicitly, we have
\begin{equation}
\partial_t \tilde{\Gamma}^i = 2 e^{4 \phi} \left[ \nabla_j \Sigma^{ij}
- \tilde{\Gamma}^i_{jk} \Sigma^{jk} - 6 \Sigma^{ij} \partial_j
\phi \right] \, .
\end{equation}

We then see that the minimal distortion condition $\nabla^j
\Sigma_{ij}=0$, and the Gamma-freezing condition $\partial_t
\tilde{\Gamma}^i=0$ are equivalent up to terms involving first spatial
derivatives of the spatial metric multiplied with the distortion
tensor itself.  In particular, all terms involving second derivatives
of the shift are identical in both cases (but not so terms with first
derivatives of the shift which appear in the distortion tensor
$\Sigma_{ij}$).  That the difference between both conditions involves
Christoffel symbols should not be surprising since the minimal
distortion condition is covariant while the Gamma-freezing condition
is not.

Just as it is the case with the lapse, we obtain parabolic and
hyperbolic shift prescriptions by making either $\partial_t \beta^i$
or $\partial^2 _t \beta^i$ proportional to $\partial_t
\tilde{\Gamma}^i$.  We call such conditions ``Gamma-driver''
conditions.  The parabolic Gamma driver condition has the form
\begin{equation}
\partial_t \beta^i = F_p \, \partial_t \tilde\Gamma^i \, ,
\label{eq:parabolicGammadriver}
\end{equation}
where $F_p$ is a positive function of space and time.  In analogy to
the discussion of the hyperbolic K-driver condition there are two
types of hyperbolic Gamma-driver conditions that we have considered,
\begin{equation}
\partial^2_t \beta^i = F \, \partial_t \tilde\Gamma^i - \eta \,
\partial_t \beta^i,
\label{eq:hyperbolicGammadriver}
\label{gamma2}
\end{equation}
or alternatively,
\begin{equation}
\partial^2_t \beta^i = F \, \partial_t \tilde\Gamma^i - 
(\eta - \frac{\partial_t F}{F}) \, \partial_t \beta^i,
\label{gamma0}
\end{equation}
where $F$ and $\eta$ are positive functions of space and time.  For
the hyperbolic Gamma-driver conditions we have found it crucial to add
a dissipation term with coefficient $\eta$ to avoid strong
oscillations in the shift.  Experience has shown that by tuning the
value of this dissipation coefficient we can manage to almost freeze
the evolution of the system at late times.

The rational behind the two almost identical choices of Gamma-driver
is the following. First note that if $F$ is independent of time, the
two choices are identical. However, we typically choose $F$ to be
proportional to $\alpha^p$, with $p$ some positive power (usually
$p=1$). Anticipating a collapsing lapse near the black hole this
implies that the term $F \partial_t \tilde\Gamma^i$ approaches zero
and the evolution of the shift tends to freeze independent of the
behaviour and numerical errors of $\partial_t \tilde\Gamma^i$.  We
implement the first choice of the Gamma-driver, (\ref{gamma2}), as

\beq \partial_t \beta^i = B^i, \qquad \partial_t B^i = F
\partial_t\tG^i -\eta B^i, \eeq

\noindent and the second choice, (\ref{gamma0}), as

\beq \partial_t \beta^i = F B^i, \qquad \partial_t B^i =
\partial_t\tG^i -\eta B^i.
\eeq

The second variant has the advantage that if $F$ approaches zero due
to the collapse of the lapse near a black hole, then
$\partial_t\beta^i$ also approaches zero and the shift freezes.  With
the first variant, on the other hand, it is only $\partial_t
B^i=\partial_t^2 \beta^i$ the one that approaches zero, which means
the shift can still evolve. Both Gamma-drivers can give stable black
hole evolutions, although the second one leads to less evolution near
the black holes.

An important point that needs to be considered when using the
hyperbolic Gamma-driver condition is that of the gauge speeds.  Just
as it happened with the lapse, the use of a hyperbolic equation for
the shift introduces new ``gauge speeds'' associated with the
propagation of the shift.  In order to get an idea of how these gauge
speeds behave, we will consider for a moment the shift
condition~(\ref{eq:hyperbolicGammadriver}) for small perturbations of
flat space (and taking $\eta$=0). From the form of $\partial_t
\tilde\Gamma^i$ given by equation (\ref{dtGdt}) we see that in such a
limit the principal part of the evolution equation for the shift
reduces to
\begin{equation}
\partial^2_t \beta^i = F \left( \delta^{jk} \partial_j \partial_k \beta^i
+ \frac{1}{3} \, \delta^{ij} \partial_j \partial_k \beta^k \right ) \, .
\end{equation}

Consider now only derivatives in a given direction, say $x$. We find
\begin{equation}
\partial^2_t \beta^i = F \left( \partial^2_x \beta^i
+ \frac{1}{3} \, \delta^{ix} \partial_x \partial_x \beta^x \right ) \, ,
\end{equation}
which implies
\begin{eqnarray}
\partial^2_t \beta^x &=& \frac{4}{3} \, F \partial^2_x \beta^x \, , \\
\partial^2_t \beta^q &=& F \partial^2_x \beta^q \qquad q \neq x \, .
\end{eqnarray}
We can then see that in regions where the spacetime is almost flat, the
longitudinal part of the shift propagates with speed $v_{\rm
long} = 2 \sqrt{F / 3}$ while the transverse part propagates with
speed $v_{\rm trans} = \sqrt{F}$. We therefore choose
\begin{equation}
F(\alpha) = \frac{3}{4} \, \alpha ,
\end{equation}
in order to have the longitudinal part of the shift propagate with the
speed of light.  The transverse part will propagate at a different
speed, but its contribution far away is typically very small.

In the next section we will turn to puncture evolutions. Both
$f(\alpha)$ and $F(\alpha)$ will be further adjusted for the presence
of punctures.


\section{Punctures}
\label{se:punctures}

So far our discussion of the BSSN formulation and the proposed gauge
conditions was quite independent of any particular choice of initial
data, except that our gauge conditions are tailored for the late time
stationarity of binary black hole mergers even though they are also
applicable in more general situations. In this Section we introduce
puncture initial data for black holes and the method of puncture
evolutions.

\subsection{Puncture initial data}
\label{se:puncinit}

Consider the three-manifold $R^3$ with one or more points
$(x_A,y_A,z_A)$ removed. These points we call punctures. The punctured
$R^3$ arises naturally in solutions to the constraints in the
Lichnerowicz-York conformal method~\cite{Lichnerowicz44,York79} for
the construction of black hole initial data. In the conformal method,
we introduce on the initial hypersurface at $t = 0$ the conformal
variables $\bar\gamma_{ij}$ and $\bar A_{ij}$ by
\bea
\gamma_{ij} &=& \psi^4_0 \bar\gamma_{ij}, \\
A_{ij} &=& \psi^{-2}_0 \bar A_{ij},
\eea
where $\psi_0$ is the conformal factor, and leave $K$
untransformed. Note that $\tA_{ij} = \psi^{-6}_0 \bar A_{ij}$ at 
$t=0$.

Consider initial data with the conformally flat metric
\beq
\gamma_{ij} = \psi^4_0 \delta_{ij}.
\eeq
Assuming that the extrinsic
curvature $K_{ij}$ vanishes, the momentum constraints
(\ref{Dconstraint}) are trivially satisfied and the Hamiltonian
constraint (\ref{Hconstraint}) reduces to
\beq \Delta^\delta \psi_0 = 0,
\label{laplacepsi}
\eeq
where $\Delta^\delta$ is the flat space Laplacian.
A particular solution to this equation is
\beq
\psi_0 = 1 + \frac{m}{2r},
\eeq
where $r^2 = x^2 + y^2 + z^2$ and we assume $r\neq 0$. This way we
have obtained initial data representing a slice of a Schwarzschild
black hole of mass $m$ in spatially isotropic coordinates on a
punctured $R^3$.  The horizon is located at $r = m/2$.  There
are two asymptotically flat regions, one for $r\rightarrow\infty$ and
a second one at $r\rightarrow0$. In fact, the metric is isometric
under $r' = m^2/(4r)$. Since (\ref{laplacepsi}) is linear in
$\psi_0$ one immediately obtains solutions for multiple black holes, for
example
\beq
\psibl = 1 + \frac{m_1}{2r_1} + \frac{m_2}{2r_2},
\eeq
where $r^2_A = (x-x_A)^2 + (y-y_A)^2 + (z-z_A)^2$ on an $R^3$ with two
punctures at $(x_A,y_A,z_A)$, $A=1,2$. These solutions were first mentioned
in~\cite{Misner57} and studied in detail by Brill and Lindquist
in~\cite{Brill63}. While no longer isometric, this initial data
contains one or two black holes depending on the separation of the
punctures, but in any case with two separate inner asymptotic regions at
the punctures. In particular, there is no physical singularity at the
punctures, but there is a coordinate singularity at each puncture if
one considers the unpunctured $R^3$.

Brill-Lindquist data can be generalized to longitudinal,
non-vanishing extrinsic curvature $K_{ij}$ for multiple black holes with
linear momentum and spin~\cite{Beig96,Brandt97b}. Here one uses the
Bowen-York extrinsic curvature,
\bea
\bar A^{ij} &=& \frac{3}{2r^2} 
(n^i P^j + n^j P^i - (\delta^{ij} - n^i n^j)) \delta_{kl} n^k P^l) 
\nonumber \\
&& + \frac{3}{r^3} (n^i \epsilon^{jkl} S_k n_l + n^j \epsilon^{ikl} S_k n_l),
\label{BYK}
\eea
and $K = 0$. The parameters $P^i$ and $S^i$ are the linear and angular
ADM momentum, and $n^i = x^i/r$ is the coordinate normal vector. The
sum of two Bowen-York terms centered at two punctures is an explicit
solution to the conformal momentum constraint with $K=0$.

The key observation for puncture initial data is that, even though
there is a coordinate singularity at each puncture, both in the
conformal factor and in the Bowen-York extrinsic curvature, we can
rewrite the Hamiltonian constraint as a regular equation on $R^3$
without any puncture points removed. This equation possesses a unique
solution $u$ that is $C^2$ at the punctures and $C^\infty$ elsewhere,
and the original Hamiltonian constraint is solved by

\beq
\label{psipuncdata}
\psi_0 = u + \psibl.
\eeq
Working on $R^3$ simplifies the numerical solution of the constraints
over methods that for example use an isometry condition at the throat
of a black hole.

Note that the puncture method for initial data can be applied using a
conformally flat metric and the Kerr extrinsic curvature \cite{Dain:2002ee},
and also to non-conformally flat initial data for multiple Kerr black
holes~\cite{Dain00,Dain01a,Dain01b}. In this paper we restrict
ourselves to the conformally flat puncture data with Bowen-York
extrinsic curvature.

\subsection{Puncture evolutions in the ADM system}
\label{se:puncevol}

In this section we want to argue that one can obtain regular
evolutions of puncture initial data without removing a region
containing the puncture coordinate singularity from the grid by, say,
an isometry condition at the throat of the black holes as
in~\cite{Anninos93b}, or through black hole
excision~\cite{Thornburg87,Seidel92a}. Evolving on $R^3$ instead of on
$R^3$ with a sphere removed and an additional boundary condition
imposed results in a significant technical simplification.

That this is possible was first noticed experimentally for a single
Schwarzschild puncture in~\cite{Anninos94c} for ADM evolutions with
singularity avoiding slicing and vanishing shift. By turning off the
isometry condition at the throat and computing everywhere including
next to the puncture, the lapse equation can still be solved for a
numerically smooth lapse with vanishing first derivative at the
puncture that collapses to zero at and around the puncture during the
evolution. The numerical grid in these simulations is staggered around
the puncture points.

In~\cite{Bruegmann97}, puncture evolutions are proposed as a general
method for the evolution of the conformally flat, longitudinal
extrinsic curvature data of orbiting and spinning black holes
discussed in Sec.~\ref{se:puncinit}. In particular, an argument is
given that {\em the punctures do not evolve by construction}.  This is
not a theorem about the regularity of the solutions, as is available
for puncture initial data~\cite{Beig96,Brandt97b}, but it is
consistent with the numerical results.

The basic idea is to examine the evolution equations and the gauge
conditions at $t = 0$ in the limit of small distance to one of the
punctures. For simplicity we move one of the punctures onto the origin
and consider the limit $r \rightarrow 0$.

In this Section we will give a detailed version of the argument
of~\cite{Bruegmann97} for the ADM equations with maximal slicing and
vanishing shift, and then discuss the BSSN equations in
Sec.~\ref{se:puncevolbssn}.  First note that for the puncture initial
data based on (\ref{BYK}) and (\ref{psipuncdata}) we have $\psibl =
O(1/r)$ and $u = O(1)$, and therefore at $t = 0$,
\bea
\psi_0 &=& O(1/r), 
\\
\psi_0^{-1} &=& O(r),  
\\
\gamma_{ij} &=& \psi_0^4 \delta_{ij} = O(1/r^4),
\\ 
K_{ij} &=& \psi_0^{-2} \bar A_{ij} = O(1/r). 
\eea
We therefore observe that the ADM equations (\ref{dgdt}) and
(\ref{dKdt}) for the evolution of the metric and the extrinsic
curvature are singular at the punctures.

The basic construction in puncture evolutions is to factor out the
{\em time-independent} conformal factor $\psibl$ (and not $\psi_0$)
given by the initial data,
\bea
\gamma_{ij} &=& \psibl^4 \tg_{ij},
\label{gammapsiBL}
\\
K_{ij} &=& \psibl^4 \tK_{ij}.
\label{KpsiBL}
\eea
The key difference to the BSSN rescaling is that puncture evolutions
involve a special rescaling that is independent of time.

Eq.~(\ref{gammapsiBL}) gives rise to a method for
accurate finite differencing of the metric. For example, for the first
partial derivative we have
\beq
\partial_k \gamma_{ij} = \psibl^4 \partial_k \tg_{ij} + 
                         \tg_{ij} \partial_k \psibl^4,
\eeq
where $\psibl^4$ and $\partial_k \psibl^4$ are given
analytically, and $\tg_{ij}$ and $\partial_k \tg_{ij}$ are assumed to
remain regular during the evolution. By staggering the puncture
between grid points one can therefore still obtain accurate
derivatives of $\gamma_{ij}$ near the puncture, and this applies to
all quantities derived from the metric and its derivatives like the
Christoffels and the Ricci tensor. In particular, there is no finite
differencing across the singularity of $1/r$ terms.

In general, we have $\psi_0 = u + \psibl$ and the analytic
derivatives of $\psi_0$ are not known, but we can still factor out
$\psibl$ as in (\ref{gammapsiBL}) and (\ref{KpsiBL}) and obtain
regular initial data,
\bea
\tg_{ij} &=& \psibl^{-4}\psi_0^4 \delta_{ij} = O(1),
\\
\tK_{ij} &=& \psibl^{-4}\psi_0^{-2} \bar A_{ij} = O(r^3).
\eea
The question is whether $\tg_{ij}$ and $\tK_{ij}$ develop a
singularity during the course of the evolution.

The ADM equations for $\tg_{ij}$ and $\tK_{ij}$ in the case of
vanishing shift are
\bea
\partial_t\tg_{ij} &=& - 2\alpha\tK_{ij},
\\
\partial_t\tK_{ij} &=& \psibl^{-4} (- D_i D_j \alpha + \alpha R_{ij})
\nonumber \\
&&  + \tg^{kl} (\tK_{ij} \tK_{kl} - 2 \tK_{ik} \tK_{jl}).
\eea
Let us examine the terms on the right hand side of the
$\partial_t\tK_{ij}$ equation. For $\tg_{ij} = O(1)$ and $\tK_{ij} =
O(r^n)$, the terms involving $\tK_{ij}$ are of order
$O(r^{2n})$. According to (\ref{Christoffel}), $\Gamma^k_{ij} =
\tG^k_{ij} + (\Gamma_{\psibl})^k_{ij}$. Assuming that $\tg_{ij}$ and
its derivatives are $O(1)$, we have $\tG^k_{ij}=O(1)$, but
$(\Gamma_{\psibl})^k_{ij} = O(1/r)$. Hence $\Gamma^k_{ij} = O(1/r)$,
and similarly $R_{ij} = O(1/r^2)$.  Finally, let us also assume that
the lapse and its derivatives are of order $O(1)$. Then $\psibl^{-4}
D_i D_j \alpha = O(r^3)$.

With these assumptions we obtain for $t=0$ that
\bea
\partial_t\tg_{ij}(0) &=& O(r^3),
\label{dtdgpunc0}
\\
\partial_t\tK_{ij}(0) &=& O(r^2),
\label{dtdKpunc0}
\eea
where the $O(r^2)$ in the last equation is contributed by the term
involving the Ricci tensor, the lapse terms are $O(r^3)$, and the
extrinsic curvature terms are $O(r^6)$. In order to study the time
evolution, we can perform one finite differencing step in time from
$t=0$ to $t=\Delta t$, for example,
\bea
\tg_{ij}(\Delta t) &=& \tg_{ij}(0) + \Delta t \partial_t\tg_{ij}(0) = O(1),
\\
\tK_{ij}(\Delta t) &=& \tK_{ij}(0) + \Delta t \partial_t\tK_{ij}(0) = O(r^2).
\eea
Note that $\tK_{ij}$ has dropped from $O(r^3)$ to $O(r^2)$. However,
it is readily checked that a second finite time step does not further
lower the order of any variable since the order of the right hand side
in the evolution equation of $\tK_{ij}$ is dominated by $\psibl^{-4}R_{ij}$.
We therefore find that 
\bea
&& \tg_{ij}(t) = O(1), \qquad \partial_t\tg_{ij}(t) = O(r^2),
\label{dtdgpunc}
\\
&& \tK_{ij}(t) = O(r^2), \qquad \partial_t\tK_{ij}(t) = O(r^2). 
\label{dtdKpunc}
\eea
This argument suggests that if the lapse $\alpha$ and its derivatives
do not introduce additional singularities at the punctures, and if
there are no singularities appearing in the spatial derivatives of the
metric (which we have not completely ruled out), then the right-hand
sides of the evolution equations for $\tg_{ij}$ and $\tK_{ij}$ vanish
at the punctures for all times. This means that $\tg_{ij}$ and
$\tK_{ij}$ should not evolve at all at the punctures for a regular
slicing and vanishing shift by construction, and the same is true for
$g_{ij}$ and $K_{ij}$.

\subsection{Puncture evolutions in the BSSN system}
\label{se:puncevolbssn}

For accurate finite differencing in the BSSN system for puncture data
we split the logarithmic conformal factor $\phi$ into a singular but
time-independent piece and an additional time-dependent contribution
$\chi$,
\beq
\phi = \chi + \ln \psibl.
\eeq
It remains to be seen whether $\chi$ and the remaining BSSN quantities
are regular throughout the evolution, i.e.\ whether the coordinate
singularity can be cleanly separated out with $\psibl$ as in the case
of ADM. To decide this issue we have to be specific about our gauge
choice. In preparation for the discussion of the gauge for puncture
evolutions in Sec.~\ref{se:gaugepuncevol} we note some
properties of the BSSN system near the punctures. 

Each of the BSSN variables has the following initial value for
puncture data at $t=0$, which we assume to evolve as indicated by the
arrows:
\bea
\label{datafalloffphi}
  \chi &=& O(1) \rightarrow O(1),
\\
  K &=& 0 \rightarrow O(r^2),
\\
  \tg_{ij} &=& O(1) \rightarrow O(1),
\\
  \tA_{ij} &=& O(r^3) \rightarrow O(r^2),
\\
  \tG^i &=& 0 \rightarrow O(r).
\label{datafalloffG}
\eea
Assume furthermore that $\alpha = O(1)$, and that the
derivatives of the $O(1)$ quantities are $O(1)$. Consider now the
following form of the evolution equations, where we have inserted our
assumptions for $\alpha$, $\tg_{ij}$, $\tA_{ij}$, and $\tG^i$, but
have kept the explicit dependence on $\beta^i$, $\phi$ and $K$:
\bea
\label{evolfalloffphi}
  \partial_t \chi &=& \lb\phi - \frac{1}{6}\alpha K,
\\
  \partial_t K &=& \lb K + O(r^4)(O(1) + O(\partial\phi)) + 
                   \frac{1}{3} \alpha K^2,
\\
  \partial_t \tg_{ij} &=& \lb\tg_{ij} + O(r^2),
\\
  \partial_t \tA_{ij} &=& \lb\tA_{ij} + K O(r^2) + O(r^4) \left( O(1) 
\right. \nonumber \\
&& \left. + O(\partial\phi) + O(\partial^2\phi) + O(\partial\phi)^2 \right),
\\
  \partial_t \tG^i &=& - \partial_j\lb\tg^{ij} + O(r^2) + 
                       O(r^2)O(\partial\phi) \nonumber \\
&& - \frac{4}{3}\alpha \tg^{ij} \partial_j K.
\label{evolfalloffG}
\eea
If these equations are to hold for all times, to be checked by
time stepping as in the last section, then we require certain
assumptions about the shift as well. In particular, each of the terms
involving $\lb$ should be of the same or higher order as the other
terms in the corresponding equation, because otherwise there could be
evolution towards lower orders in $r$.  In particular, even assuming
$\alpha$ and $K$ are regular, we have to examine the behaviour of $\lb
\phi$ at the puncture before we can conclude that $\chi$ remains
regular.

Let us assume that $\chi$ and its derivatives are regular, so that
\beq
  \phi = O(\ln r), \quad \partial_i\phi = O(1/r), \quad 
  \partial_i\partial_j \phi = O(1/r^2). 
\eeq
If furthermore $K = O(r^2)$, and if the shift terms are of
sufficiently high order, then the right hand sides of
(\ref{evolfalloffphi})-(\ref{evolfalloffG}) are at least $O(r)$. In
this case, the order of each equation is such that the corresponding
orders in (\ref{datafalloffphi})-(\ref{datafalloffG}) are maintained.
In Sec.~\ref{se:gaugepuncevol} we show that these assumptions can
indeed be met by a proper gauge choice, and hence we arrive at the
statement that in this case the punctures do not evolve by
construction.


\section{Gauge conditions and puncture evolutions}
\label{se:gaugepuncevol}

The main question is whether there are lapse and shift conditions that
behave appropriately for puncture evolutions.  We will show that this
is indeed the case without the need to introduce special boundary
conditions at the punctures. What is required is an appropriately
regularized implementation of our gauge conditions and a choice of
initial data for lapse and shift such that the punctures do not
evolve.

\subsection{Lapse for puncture evolutions}
\label{se:lapsepuncevol}

Consider maximal slicing, which is
implemented by choosing $K=0$ at $t=0$ and determining the lapse from
the elliptic equation resulting from $\partial_t K = 0$, which for
vanishing shift is
\beq
   \Delta \alpha = \alpha K_{ij} K^{ij}.
\label{maximalslicing}
\eeq
As discussed in \cite{Bruegmann97}, for $g_{ij} = \psibl^4\tg_{ij}$,
\beq
   \Delta\alpha = \psibl^{-4} \Delta^{\tg}
   \alpha - \delta^{ij}\Gamma^k_{ij}\partial_k\alpha,
\label{psideldel}
\eeq
so the principal part is degenerating to zero as $O(r^4)$. To
avoid numerical problems we therefore multiply (\ref{maximalslicing})
by $\psibl^4$, which normalizes the principal part but leaves a
$O(1/r)$ term since $\Gamma^k_{ij} = O(1/r)$: \beq \Delta^{\tg} \alpha
- O(1/r) \partial_k\alpha = O(r^6) \alpha.  \eeq It turns out that
standard numerical methods to solve this elliptic equation will find a
regular solution for $\alpha$ for which $\partial_k\alpha$ vanishes
sufficiently rapidly so that $O(1/r) \partial_k\alpha$ is zero at the
puncture. Therefore, maximal slicing and vanishing shift lead to a
sufficiently regular lapse such that indeed the right-hand sides of
the ADM evolution equations for $\tg_{ij}$ and $\tK_{ij}$ vanish at
the punctures.

\begin{figure}
\epsfxsize=85mm \epsfysize=80mm \epsfbox{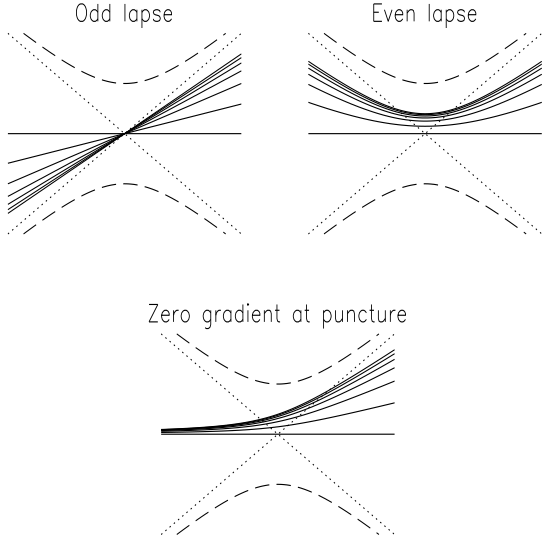} 
\caption{Schematic representation on the Kruskal diagram of the effect
  of the different boundary conditions on the slices obtained. The
  first panel shows the case of an odd lapse at the throat, the
  second panel the case of an even lapse at the throat, and the last
  panel the case of a lapse with zero gradient at the puncture.  The
  dashed lines show the singularities and the dotted lines the
  event horizon. }
\label{fig:slices}
\end{figure}

Effectively, maximal slicing implements the condition that the lapse
has a vanishing gradient at the puncture.  Notice that this condition
is very different to an isometry-type condition, where the lapse would
be forced to be $-1$ at the puncture.  Fig.~\ref{fig:slices} shows
a schematic representation in the Kruskal diagram of the type of
slices one obtains in the case of a single Schwarzschild BH when using
three different boundary conditions for the lapse (while keeping the
same interior slicing condition): odd at the throat, even at the
throat and zero gradient at the puncture.  When looking at these plots
it is important to keep in mind that the puncture corresponds to a
compactification of the second asymptotically flat region, and is at
an infinite distance to the left of the plots. Notice that, in all
three cases, far away on the right hand side of the plot the slices
approach the Schwarzschild slices (in fact, if we use maximal slicing
and ask for the lapse to be odd we recover the Schwarzschild slices
everywhere).  Also, in the case with an odd lapse the slices do not
penetrate the horizon, but in the other two cases they do.

Since maximal slicing is computationally expensive, we often use 1+log
slicing that mimics the behaviour of the maximal lapse in that it also
is singularity avoiding and the lapse drops to zero when the physical
singularity approaches.  Analytically, however, the 1+log lapse does
not necessarily drop to zero at the puncture. Starting with $\alpha =
1$ and $K=0$ everywhere at $t=0$, we see from the evolution equations
for the lapse and for $K$,
\bea
\partial_t \alpha &=& - \alpha^2 f(\alpha) (K-K_0),
\\
\partial_t K  &=& \beta^i\partial_i K + \frac{1}{3}\alpha K^2 + O(r^3),
\eea
that neither $\alpha$ nor $K$ evolve at the puncture if the shift or
the derivative of $K$ vanishes at the puncture.  That means that the
lapse will remain $1$ at the puncture, and the inner asymptotically
flat region will evolve. Numerically, one may expect this to be
problematic if there is not sufficient resolution, as will be normally
the case. The code can become unstable near the puncture, and even if
it remains stable the event horizon may not prevent numerical
inaccuracies to evolve into the outer regions. While this may happen,
in practice the 1+log lapse does collapse, apparently precisely
because of a lack of resolution, and the code remains stable. Even in
this case we typically obtain convergence in the outer regions where
the lapse did not collapse.

In this paper we experiment with 1+log slicing with $f(\alpha) =
2/\alpha$ replaced by $f(\alpha,\psibl) = 2\psibl^m/\alpha$, so that

\beq
  \partial_t \alpha = - 2 \alpha \psibl^m (K - K_0),
\eeq
i.e.\ we have introduced a factor that for $m>0$ can drive the lapse
to zero at the puncture.  For $m=0$ we obtain standard 1+log
slicing. A natural choice is $m = 4$ since then the singularity in
$\psibl^4$ exactly matches the degeneracy of the principal part of the
second order wave equation associated with the lapse evolution, see
(\ref{wavelapse}) and (\ref{psideldel}). In particular, for $m=4$ the
wave speed is regular at the puncture.

Both choices of 1+log slicing with initial lapse equal to unity have
been found to lead to stable evolutions of black holes with a lapse
that satisfies the regularity conditions assumed in the previous
sections.  Another approach to obtain a vanishing lapse at the
puncture is to start with a different initial lapse, for example

\beq 
\alpha(t=0) = \psibl^{-2} = O(r^2), 
\label{initiallapse}
\eeq
so that the lapse is zero at the puncture initially and there is no
evolution due to a non-vanishing lapse at the puncture. The power $-2$
is chosen so that the initial lapse has the same limit for
$r\rightarrow\infty$ as the lapse $\alpha_{isotropic} =
(1-\frac{m}{2r})/(1+\frac{m}{2r})$ of the static Schwarzschild metric
in isotropic coordinates. In practice, we have found that with such an
initial lapse there sometimes is too much evolution in the still
poorly resolved region between the puncture and the horizon, which is
why we do not use this option routinely. Instead of guessing an
initial lapse that minimizes the amount of initial evolution one
should use the lapse (and shift) derived from a quasi-equilibrium
thin-sandwich puncture initial data set, which however is currently
not available.


\subsection{Shift for puncture evolutions}
\label{se:shiftpuncevol}

For long term stable evolutions, we want to construct a shift
condition that counters slice-stretching. However, for arbitrary
non-vanishing shift, equations
(\ref{evolfalloffphi}-\ref{evolfalloffG}) show that the punctures will
evolve. It is possible to have the punctures move
across the grid because of a non-vanishing shift. One problem would be
the numerical treatment of the coordinate singularity at the
punctures, which so far was based on analytic derivatives of the
time-independent conformal factor $\psibl$. While a solution of
this problem may be possible, we focus here on finding a shift
condition that counters slice-stretching while simultaneously
satisfying a fall-off condition for the shift such that the punctures do
not evolve at all when using the BSSN equations.  

As a first step it is instructive to consider $\beta^i = r n^i = x^i =
O(r)$ (with $n^i$ a radial unit vector) near the puncture. In this
case several terms in the Lie derivatives cancel exactly and we have

\bea
{\cal L}_{rn} \tg_{ij} &=& x^k\partial_k \tg_{ij}, \\
{\cal L}_{rn} \tA_{ij} &=& x^k\partial_k \tA_{ij}, \\
{\cal L}_{rn} K &=& x^k\partial_k K.
\eea

However,
\bea {\cal L}_{rn} \phi &=&
x^k\partial_k \chi + x^k\partial_k \ln\psibl + \frac{1}{2} = O(1),
\eea
so $\chi$ will evolve without a special combination of $\chi$,
$K$, and $\alpha$.  Furthermore, the Lie derivative will not be as
simple if the shift is not exactly spherically symmetric.

We therefore turn to 
\beq
  \beta^i = O(r^3).
\eeq  
This happens to be the necessary condition (assuming integer powers
of $r$) for the norm of the shift in the non-conformal metric to be
zero at the puncture

\beq
  \gamma_{ij} \beta^i\beta^j = O(1/r^4) \delta_{ij} \beta^i\beta^j.
\eeq
With $\beta^i = r^3 n^i = r^2 x^i$ we now have
\bea
{\cal L}_{r^3n} \tg_{ij} &=& O(r^2) (x^k\partial_k \tg_{ij} +
\tg_{ij}) = O(r^2),
\\
{\cal L}_{r^3n} \phi &=& O(r^2),
\eea
All other Lie derivative terms also turn out to be of order $O(r^2)$.
Finally, the shift derivatives in the evolution equation for $\tG^i$ are
\beq
\partial_j\lb\tg^{ij} = \partial_j O(r^2) = O(r). 
\eeq
In this sense the evolution of $\tG^i$ poses the strictest condition
on the fall-off of the shift.

The question remains how we guarantee the $O(r^3)$ fall-off in the
actual shift condition. We can enforce such a fall-off by choosing 
\beq
\beta^i(t=0) = 0, 
\eeq 
and by changing the coefficient $F$ in the hyperbolic Gamma-driver to
\beq 
F(\alpha,\psibl) = \frac{3}{4} \, \alpha \,
{\psibl^{-n}} = O(r^n), 
\eeq 
where we typically choose $n = 2$ or $4$. Note that the two versions
of the Gamma-driver differ by the term 
$\partial_tF/F = \partial_t \alpha/\alpha$, which in the case of 1+log
slicing equals $-2\psibl^m(K-K_0)$.
Let us ignore the diffusion term.
If the shift has evolved into $\beta^i = O(r^3)$, then 
$\partial_t\tG^i = O(r)$. With $n \geq 2$ we have 
\beq 
\partial_t^2 \beta^i = O(r^3).  
\eeq 
In fact, we have found that in the case of evolutions of
just one black hole (distorted or not), changing $F$ is not really
necessary since there is enough symmetry in the problem to guarantee
that the shift will have the correct fall-off at the puncture even
without introducing the extra factor of $\psi^{-n}_{BL}$ into $F$.
When dealing with two black holes, however, this is no longer the case
and the factor $\psi^{-n}_{BL}$ is required to stop the shift from
evolving at the punctures.

Anticipating the numerical results, let us point out that due to a
lack of numerical resolution the shift often looks like $O(r)$ even
though we have chosen $n = 2$ or $4$. Somewhat surprisingly, even in
these cases the gauge conditions are able to approximately freeze the
evolution, for which we do not yet have a good analytic explanation.

\subsection{Vanishing of the shift at the punctures}

Combining our choice of puncture initial data with the lapse and shift
conditions above, we have the expectation that the BSSN
variables will not evolve at the punctures.  This is a natural
situation considering that the punctures represent an asymptotically
flat infinity, and that there is no linear momentum at the inner
infinities. Performing the transformation $r \rightarrow 1/r$ for
puncture data with the Bowen-York extrinsic curvature defined in
(\ref{BYK}) shows that the $1/r^3$ spin terms are mapped to $1/r^3$
terms, but the $1/r^2$ linear momentum terms are mapped to $1/r^4$
terms and there are no $1/r^2$ terms, and therefore there is no linear
momentum at the inner infinity.  In other words, viewed from the other
asymptotic end, the black hole does not move in the data we use.

One can add a $1/r^4$ term to $\bar A_{ij}$ to make the holes move in
the inner ends, but then the puncture initial data construction and
the puncture evolutions on $R^3$ fail for a lack of regularity at the
punctures. In general, with a different choice of extrinsic curvature
that does not satisfy the fall-off conditions of the Bowen-York data
(\ref{BYK}), there can be non-trivial or even singular evolution at
the punctures in both the ADM and the BSSN systems.

In summary, our puncture initial data corresponds to two black holes
which are momentarily at rest at their inner asymptotic ends. For a
given coordinate system the black holes could start moving if there is
a non-vanishing shift at the punctures, but we explicitly construct a
vanishing shift at the punctures. The main consequence for puncture
data of orbiting and colliding black holes is that by construction the
inner asymptotic ends of the black hole will not move in our
coordinate system, i.e.\ the punctures remain glued to the grid. That
still allows for general dynamics around the punctures, which shows in
the evolution of the metric and extrinsic curvature. For example, the
apparent horizon can easily grow, drift and change shape, but it {\em
can not} cross over the punctures for geometrical reasons, since the
apparent horizon area would become infinite before they could do
that. For orbiting black holes, since the punctures do not move by
construction, it seems natural to combine the method of puncture
evolutions with a corotating coordinate system to minimize the
evolution of the metric data. We leave this option for future work.


\section{Numerics}
\label{se:numerics}

The numerical time integration in our code uses an iterative
Crank-Nicholson scheme with 3 iterations, see
e.g.~\cite{Alcubierre99d}.  Derivatives are represented by second
order finite differences on a Cartesian grid. We use standard centered
difference stencils for all terms, except in the advection terms
involving the shift vector (terms involving $\beta^i
\partial_i$).  For these terms we use second order upwind along the
shift direction.  We have found the use of an upwind scheme in such
advection-type terms crucial for the stability of our code.  Notice
that this is the only place in our implementation where any
information about causality is used (i.e.\ the direction of the tilt in
the light cones).

\subsection{Outer boundary condition}

At the outer boundary we use a radiation (Sommerfeld) boundary
condition.  We start from the assumption that near the boundary all
fields behave as spherical waves, namely we impose the condition
\begin{equation}
f = f_0 + u(r-vt)/r .
\label{eq:radiation}
\end{equation}
Where $f_0$ is the asymptotic value of a given dynamical variable
(typically 1 for the lapse and diagonal metric components, and zero
for everything else), and $v$ is some wave speed.  If our
boundary is sufficiently far away one can safely assume that the speed
of light is 1, so $v=1$ for most fields.  However, the gauge variables
can easily propagate with a different speed implying a different value
of $v$.

In practice, we do not use the boundary condition~(\ref{eq:radiation})
as it stands, but rather we use it in differential form:
\begin{equation}
\partial_t f + v \partial_r f - v \, (f - f_0)/r = 0 .
\end{equation}
Since our code is written in Cartesian coordinates, we transform the
last condition to
\begin{equation}
\frac{x_i}{r} \, \partial_t f + v \partial_i f + \frac{v x_i}{r^2} \,
\left( f - f_0 \right) = 0 .
\end{equation}
We finite difference this condition consistently to second order in
both space and time and apply it to all dynamic variables (with
possible different values of $f_0$ and $v$) at all boundaries.

There is a final subtlety in our boundary treatment.  Wave propagation
is not the only reason why fields evolve near a boundary.  Simple
infall of the coordinate observers will cause some small evolution as
well, and such evolution is poorly modeled by a propagating wave. This
is particularly important at early times, when the radiative boundary
condition introduces a bad transient effect. In order to minimize the
error at our boundaries introduced by such non-wavelike evolution, we
allow for boundary behavior of the form:
\begin{equation}
f = f_0 + u(r-vt)/r + h(t)/r^n ,
\label{eq:radpower}
\end{equation}
with $h$ a function of $t$ alone and $n$ some unknown power. This
leads to the differential equation
\bea \partial_t f + v \partial_r f
- \frac{v}{r} \, (f - f_0) &=&
\frac{v h(t)}{r^{n+1}} \left( 1 - n v \right)
+ \frac{h'(t)}{r^n} \nonumber \\
&\simeq& \frac{h'(t)}{r^n} \qquad \mathrm{for\ large\ } r ,
\eea
or in Cartesian coordinates
\begin{equation}
\frac{x_i}{r} \, \partial_t f + v \partial_i f + \frac{v x_i}{r^2} \,
\left( f - f_0 \right) \simeq \frac{x_i h'(t)}{r^{n+1}} .
\end{equation}

This expression still contains the unknown function $h'(t)$. Having
chosen a value of $n$, one can evaluate the above expression one point
away from the boundary to solve for $h'(t)$, and then use this value
at the boundary itself.  Empirically, we have found that taking $n=3$
almost completely eliminates the bad transient caused by the radiative
boundary condition on its own.

\subsection{Fish-eye transformation}

Setting up a reasonable numerical simulation, there is always the
conflicting interest of having the boundary as far out as possible and
having as good resolution as possible. With limited numerical
resources it is almost never possible to obtain both at the same
time. One way to stretch limited resources as far as possible, is to
introduce a radial coordinate transformation that decreases the
resolution with distance. Such coordinate transformations can also be
applied to 3D Cartesian grids, see the ``fish-eye transformation''
in \cite{Baker00b,Baker:2001sf}.

In order to make the outer boundary conditions as
simple as possible, we would like for the resolution to be constant
at the location of the outer boundaries. That is, we want constant high
resolution in the region containing the black holes, then we want a
region where the resolution decreases with distance and finally we want
a region (containing the outer boundaries) with constant low resolution.
Denoting the physical radius by $r$ and the coordinate radius by $r_{c}$,
the previous requirements can be met with the following radial coordinate
transformation
\begin{eqnarray}
r & = & a r_{c} + ( 1 - a ) \frac{s}{2 \tanh (\frac{r_{0}}{s})} \left [ 
          \ln \left ( \cosh\frac{r_{c}+r_{0}}{s} \right ) \right. \nonumber \\
  &   & \left. - \ln \left ( \cosh\frac{r_{c}-r_{0}}{s} \right ) \right ],
\label{eq:trans_fe}
\end{eqnarray}
where $a$ is a parameter specifying the scale factor in grid spacing, 
$r_{0}$ is the radius of the transition region and $s$ is the width
of the transition region.

By differentiating $r$ in equation~({\ref{eq:trans_fe}}) with respect
to $r_{c}$ we find that $dr/dr_{c}=1$ for $r_{c}=0$ and $dr/dr_{c}=a$
for $r_{c}\rightarrow\infty$ as required. Note that the radial $r$
coordinate is mapped to $a r_c$ plus a non-vanishing constant, and
therefore the Jacobian of this transformation does not correspond to
just a simple rescaling of radial resolution. The
transformation~(\ref{eq:trans_fe}) we refer to as the ``transition
fish-eye''.

An important point to keep in mind when using a fish-eye
transformation is the fact that both the asymptotic values of metric
components and the physical speed of light (and gauge speeds) will be
affected by the transformation.  This means that special care should
be taken when applying boundary conditions.


\section{Applications}
\label{se:applications}

In the numerical application of our method we focus on establishing
the basic validity of the puncture evolutions with the hyperbolic
shift. We consider evolutions of the spherically symmetric
Schwarzschild spacetime, of a single distorted black hole, and of the
head-on collision of two Brill-Lindquist punctures.  We will report on
orbiting binary systems elsewhere.


\begin{figure}[t]
\vspace*{14mm}
   \epsfxsize=85mm \epsfysize=60mm \epsfbox{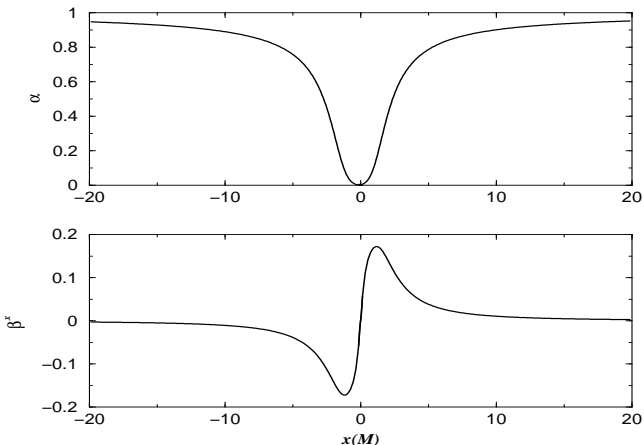} 
\caption{Schwarzschild black hole evolved for $t = 1000M$.
Shown are lapse $\alpha$ and shift component $\beta^x$ along the
x-axis, which are (anti-)symmetric about $x = 0$. By that time lapse
and shift are approximately static.  The lapse has collapsed to zero
at the puncture and approaches one in the outer region. The shift
crosses zero at the puncture, pointing away from the puncture and
thereby halting the infall of points towards the puncture.  }
\label{fig:schw_alp_beta}
\end{figure}

\begin{figure}[t]
\vspace*{9mm}
   \epsfxsize=85mm \epsfysize=60mm \epsfbox{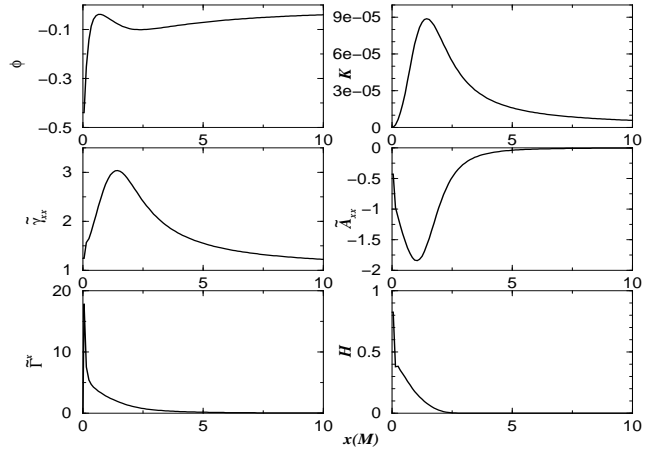} 
\caption{Schwarzschild black hole evolved for $t = 1000M$.
Shown are the BSSN variables $\phi$, $K$, $\tg_{xx}$, $\tA_{xx}$, and
$\tG^x$ along the x-axis, and also the Hamiltonian constraint $H$.
}
\label{fig:schw_g0}
\end{figure}

\subsection{Evolution of a single Schwarzschild puncture}

For the evolution of Schwarzschild we use the Cartoon method of
\cite{Alcubierre99a} for implementing axisymmetric systems with 3D
Cartesian finite differencing stencils. Choosing the z-axis as the
axis of symmetry, we evolve a 3D Cartesian slab with just 5 points in
the y-direction. On the $y=0$ plane standard 3D stencils are computed,
and the data at the points with $y\ne 0$ are obtained by
interpolation in the x-direction in the $y=0$ plane and by tensor rotation
about the z-axis. For Schwarzschild we also use the reflection symmetry
in the $z=0$ plane.

We choose the Schwarzschild puncture data of Sec.~\ref{se:puncinit}
with $m=1.0M$ and the apparent horizon at $r=0.5M$.
As we have discussed, there are several choices for the gauge
conditions. For the Schwarzschild puncture, we initialize lapse and shift to
$\alpha = 1$ and $\beta^i = 0$. We
consider 1+log slicing, (\ref{eq:hypKdriver}), and the
hyperbolic shift, (\ref{gamma0}), with the specific choice of
\beq
	f = 2 \alpha^{-1} \psibl^4, 
	\;\; F = \frac{3}{4} \alpha\psibl^{-2},
	\;\; \eta = 2.0/M.
\label{schwgauge1}
\eeq
In Fig.~\ref{fig:schw_alp_beta} we show lapse and shift for an
evolution with 201 points in x- and z-direction, starting at the
staggered point at the origin and extending to about $20M$ with a grid
spacing of $0.1M$. We plot the data after an evolution of $t = 1000M$,
which corresponds to 40000 time steps with a Courant factor of $0.25$.

\begin{figure}
\vspace*{8mm}
   \epsfxsize=85mm \epsfysize=60mm \epsfbox{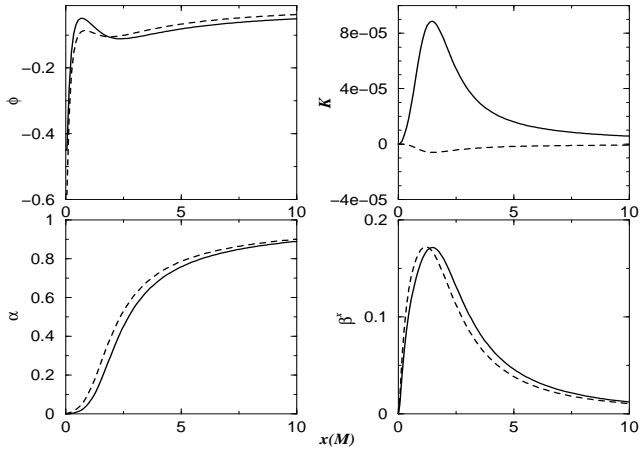} 
\caption{Schwarzschild black hole evolved for $t = 1000M$.
Shown is a comparison along the x-axis between two versions of the hyperbolic
Gamma-driver for the shift, Eq.~(\ref{gamma2}) (dashed line) and
Eq.~(\ref{gamma0}) (solid line). 
}
\label{fig:schw_g0_g2}
\end{figure}

\begin{figure}
\vspace*{5mm}
   \epsfxsize=85mm \epsfysize=60mm \epsfbox{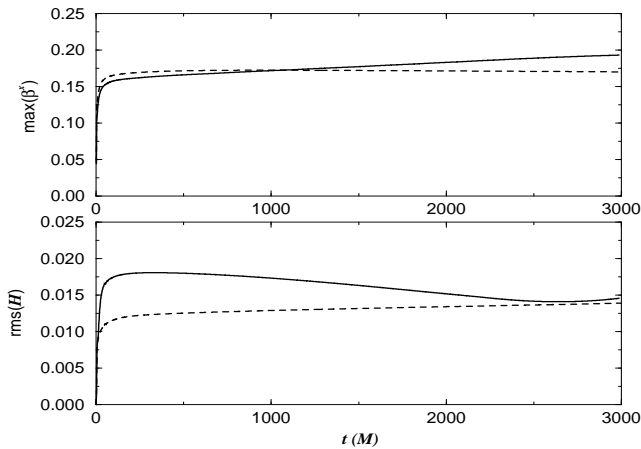} 
\caption{Schwarzschild black hole evolved for $t = 3000M$.
Shown are the maximum of the shift and the root-mean-square value of
the Hamiltonian constraint as a function of time, again for two
versions of the hyperbolic Gamma-driver for the shift,
Eq.~(\ref{gamma2}) (dashed line) and Eq.~(\ref{gamma0}) (solid line),
with diffusion parameter $\eta = 2.0/M$ and $\eta = 2.8/M$,
respectively.  After a short time interval during which lapse and
shift adjust themselves dynamically, the evolution slows down
significantly.  }
\label{fig:schw_beta_ham_t}
\end{figure}

Lapse and shift show the characteristic feature of puncture
evolutions. Both lapse and shift are zero at the puncture, indicating
that there is no evolution at the inner asymptotically flat end of the
black hole. The lapse approaches one in the outer region, while the
shift points outward from the puncture and approaches zero in the
outer region. The shift counters the infall of points towards the
puncture, thereby stopping the slice stretching.

Fig.~\ref{fig:schw_g0} shows various other quantities of the same
Schwarzschild run at $t=1000M$. Note the behaviour near the puncture,
which at this resolution appears to be regular but is not sufficiently
resolved.

In Fig.~\ref{fig:schw_g0_g2} we compare data from this run with a run
for identical parameters except that instead of Eq.~(\ref{gamma0}) we
use Eq.~(\ref{gamma2}) with $\eta = 2.8/M$ for the shift. The
differences are quite small in the case of this Schwarzschild run.

In Fig.~\ref{fig:schw_beta_ham_t} we show the maximum of the shift and
the root-mean-square value of the Hamiltonian constraint as a function
of time. After a short time interval of less than $100M$ (recall that
previous runs with vanishing shift lasted only to about $30$-$40M$!),
evolution is approximately frozen for more than $3000M$. The observed
drift in various quantities is crucially affected by the value of
$\eta$ that determines the diffusion in the hyperbolic
Gamma-driver. In Fig.~\ref{fig:schw_beta_ham_t} we compare again the
two versions of the Gamma-driver, and note that two different values
of $\eta$, $2.0/M$ and $2.8/M$, are used to obtain long term stability.
It is a matter of experimentation to find a suitable value of $\eta$
in dependence on the various parameters in the run. Runs may crash
before $100M$ for a bad choice of $\eta$.  On the other hand, once
determined for a particular initial data set and set of grid
parameters, we found that the runs were rather robust under small
variations. It would be useful to have a dynamic determination and
adaptation of $\eta$, but this is currently not available.

Having established the basic features and the stability of our gauge
choice, we want to study convergence for Schwarzschild. A crucial
issue is whether we obtain convergence near the puncture.  We choose
three grid sizes and resolutions: 201 points in both the x- and
z-directions for a resolution of $0.025M$, 401 points at $0.0125M$,
and 801 points at $0.00625M$.  With a Courant factor of $0.25$ in the
BSSN evolution scheme it takes 160, 320, and 640 iterations,
respectively, for an evolution time of $1M$. The outer boundary is at
about $5M$. We choose the same gauge as in (\ref{schwgauge1}), except
that in $F$ we use $\psibl^{-4}$ instead of $\psibl^{-2}$ for a
broader profile of the shift near the puncture.

\begin{figure}
\vspace*{-12mm}
   \epsfxsize=85mm \epsfysize=60mm \epsfbox{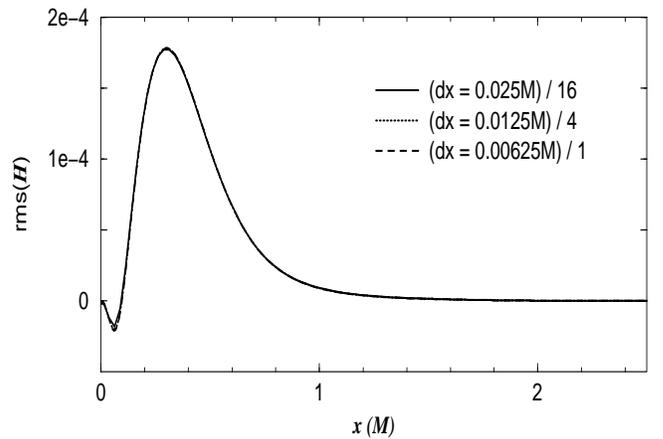} 
\caption{Schwarzschild black hole evolved to $t=5M$ at three high resolutions,
demonstrating second order convergence at the puncture.}
\label{fig:schw_ham}
\end{figure}

Fig.~\ref{fig:schw_ham} shows the Hamiltonian constraint along the
x-axis near the single Schwarzschild puncture at the three
resolutions, rescaled by the corresponding factors expected for second
order convergence. The coincidence of the three lines indicates clean
second order convergence.
Therefore, for such high resolutions the BSSN system exhibits the
expected regular and convergent behaviour near the punctures. 
It is remarkable that even at a four times coarser
resolution of $0.1M$ the evolutions remain stable. 

\begin{figure}[t]
\vspace*{-5mm}
\epsfxsize=85mm \epsfysize=60mm 
\epsfbox{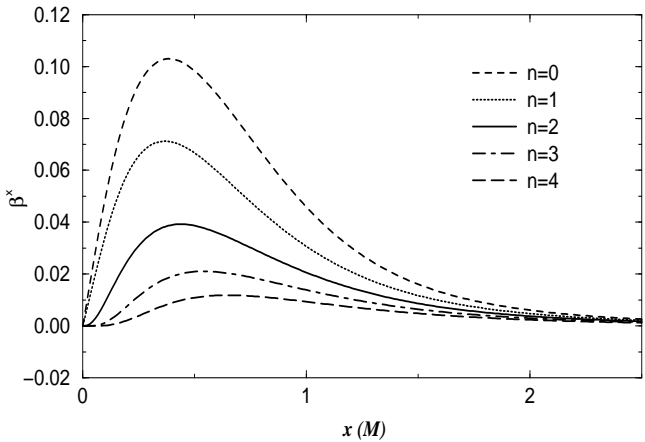}
\caption{Schwarzschild black hole evolved for $t = 1M$.
Shown is the effect of varying the power $n$ in $\psibl^{-n}$ in the shift
equation for $\beta^x$ along the x-axis.
}
\label{fig:schw_psi_n}
\end{figure}

Note in particular that the shift in Fig.~\ref{fig:schw_alp_beta}
seems to be linear at the puncture, in contrast with the expected
$O(r^3)$ behaviour. Fig.~\ref{fig:schw_psi_n} shows the effect of
different powers of $\psibl$ in the shift equation for the grid
parameters of the medium resolution run of the convergence test. 
We use the shift equation (\ref{gamma0}), and 
\beq 
F = \frac{3}{4} \alpha\psibl^{-n}, \;\; \eta = 2.0/M,
\label{schwgauge2}
\eeq
with different values for $n$. Fig.~\ref{fig:schw_psi_n} shows the
shift for $t = 1M$. Note the resolution that is required to make the
$O(r^3)$ behaviour visible for $n\geq2$. By $t = 10M$, the shift for
$n = 2$ is no longer completely resolved at the puncture with a grid
spacing of $0.0125M$, but as we have seen, even at coarser resolutions
the approximate $O(r)$ behaviour of the shift at the puncture allows
stable evolutions.

\subsection{Evolution of a single, distorted black hole}

The second application we present is that of a distorted BH. Referring
to~\cite{Brandt97c}, we choose a distortion parameter $Q_0=0.5$,
position $\eta_0=0$, and width $\sigma=1$.  The ADM mass of this
system is $M=1.83$.  Such data has been previously evolved in 2D and
in 3D using excision. Here we discuss a 3D puncture evolution with
octant symmetry, $129^3$ points and a grid spacing of $0.1M =
0.183$. The outer boundary is at about $12.8M$. For the gauge we use
1+log slicing with the initial lapse not unity but given by
(\ref{initiallapse}) and the hyperbolic shift condition (\ref{gamma0})
with
\beq
	f = 2 \alpha^{-1} \psibl^4, 
	\; F = \frac{3}{4} \alpha\psibl^{-2},
	\; \eta = 1.25/M \approx 0.68.
\label{dbhgauge}
\eeq

\begin{figure}[t]
\vspace*{5mm}
\epsfxsize=85mm \epsfysize=80mm \epsfbox{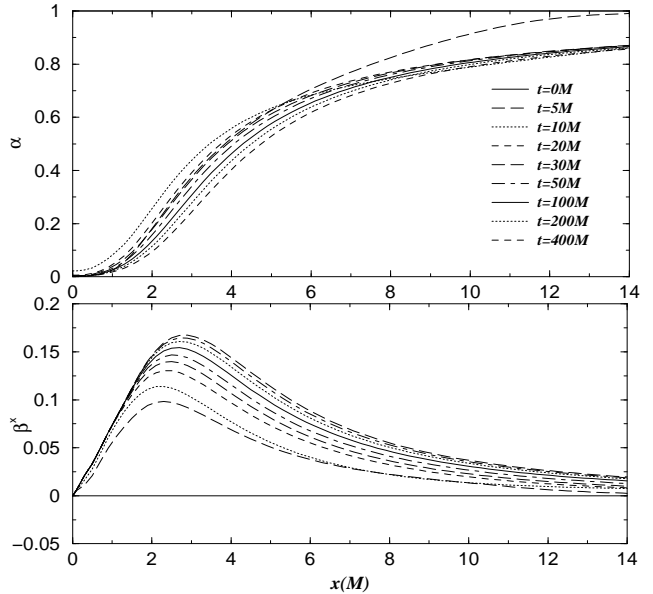} 
\caption{Lapse and shift for the evolution of a single distorted 
BH. After around $20M$, the evolution of lapse and shift 
slows down significantly (note the time labels). The approach to the
final profile in lapse and shift is not monotonic.
}
\label{gauge-distort}
\end{figure}

In Fig.~\ref{gauge-distort}, we show the evolution of the lapse and
the shift component $\beta^x$ along the x-axis. Note that the shift,
although vanishing initially, develops the needed profile simply
through its evolution equation, without any special initial condition.
After a short while, the evolution effectively freezes, allowing the
waves to propagate on an effectively fixed BH background, just as one
would like.

\begin{figure}
\epsfxsize=85mm \epsfysize=80mm \epsfbox{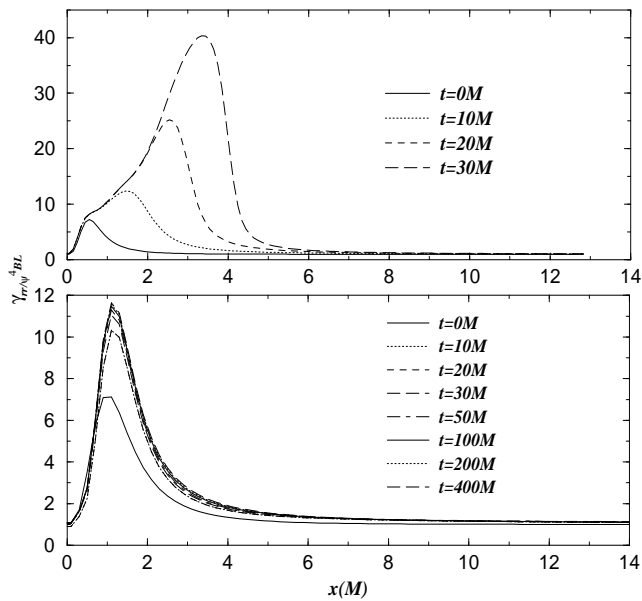} 
\caption{The evolution of the radial metric function
  $\gamma_{rr}/\psibl^{4}$ for a distorted BH along the $x-$axis.
  The upper panel shows the slice stretching in the metric for
  singularity avoiding slicing with vanishing shift, while the lower
  panel shows the metric for the new gauge conditions.  Without shift
  the metric grows out of control after $t = 40M$, while with the new
  shift condition a peak begins to form initially but later almost
  freezes as lapse and shift drive the BH into an essentially static
  configuration (note the time labels).
}
\label{grr-distort}
\end{figure}

For comparison, we show in Fig.~\ref{grr-distort} the evolution of the
radial component of the metric, $\gamma_{rr}/\psibl^4$, for
the new gauge condition (lower panel) and for a singularity avoiding
slicing run with 1+log slicing and vanishing shift (upper panel).
For 1+log slicing and vanishing shift we see the well-known slice
stretching effect. With the new gauge evolution is slowed
significantly at late times. The peak of the metric near $x = 0.5M$
grows to about 12 by time $t=20M$ and does not grow significantly after
that until $t = 400M$ (lower panel), while for vanishing shift already at $t
= 30M$ the peak in the metric has reached 40 without any sign of
slowing growth (upper panel).

\begin{figure}[t]
\epsfxsize=85mm \epsfysize=80mm \epsfbox{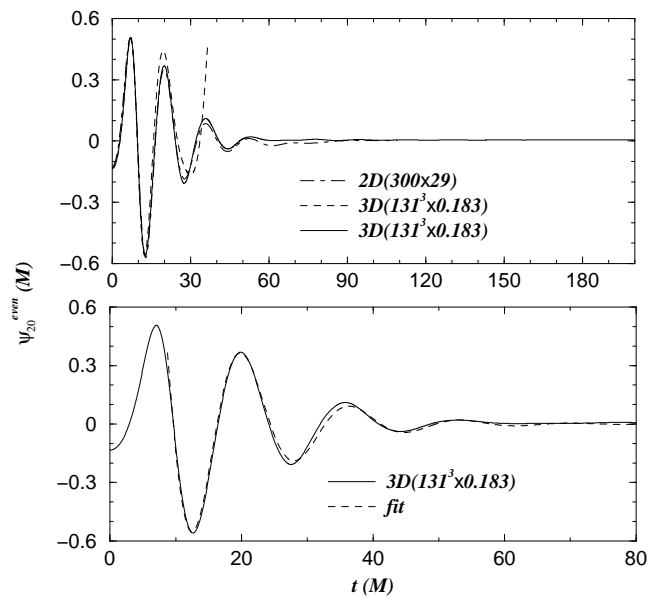} 
\caption{The solid line shows the $\ell=2, m=0$ waveform extracted at
a radius of $5.45M$ for the even-parity distorted BH described in the
text, while the two dashed lines show the result of the same
simulation carried out in the 2D and 3D code with vanishing shift. The
2D code crashes at around $t=100M$ and the 3D code crashes around
$t=40M$.  The lower panel shows a fit for the time interval from
$t=9M$ to $t=80M$ to the two lowest quasi normal modes of the BH for
the new gauge conditions in 3D, confirming that the ring-down of the
distorted BH is simulated accurately.}
\label{wave-distort}
\end{figure}

For the new gauge we expect that we can reliably extract the waveform
for the ring-down, and this is indeed the case as shown in
Fig.~\ref{wave-distort}.

\subsection{Head-on collision of two Brill-Lindquist punctures}

The third application we present is that of a head on collision of two
Brill-Lindquist BH's. The parameters for these simulations are
$m_{1}=m_{2}=0.5M$, $\mathbf{C}$$_{1}=\{0,1.1515M,0\}$,
$\mathbf{C}$$_{2}=\{0,-1.1515M,0\}$, where $m_{1}$ and $m_{2}$ are the
masses of the BH's and $\mathbf{C}$$_{1}$ and $\mathbf{C}$$_{2}$ are
the locations of the two punctures. These parameters correspond to an
initial separation of the BH's equal to that of an approximate ISCO
configuration as determined in~\cite{Baumgarte00a}. Such data has been
previously evolved without shift with the Lazarus technique that
combines short, fully numerical evolutions with a close limit
approximation for the wave extraction~\cite{Baker00b}
(see~\cite{Baker:2002qf} for runs starting at larger separation).

We present two types of runs for the head-on collision starting at the
approximate ISCO separation. In the first type we use 1+log slicing
and the hyperbolic Gamma-driver (\ref{gamma2}) with
\beq
	f = 2 \alpha^{-1}, 
	\;\; F = \frac{3}{4} \alpha\psibl^{-4},
	\;\; \eta = 2.8/M.
\label{bbh0gauge}
\eeq
with an initial lapse equal to one and an initial shift equal to
zero.  We also use the transition fish-eye with parameters $a = 3$, $s
= 1.2M$ and $r_{0}=5.5M$. This places the outer boundary at a distance
of $25.8M$ with central resolutions $0.128M$, $0.064M$ and $0.032M$ and
gridsizes $96^{3}$, $192^{3}$ and $384^{3}$ respectively in octant
mode.

\begin{figure}[t]
\epsfxsize=85mm \epsfysize=60mm \epsfbox{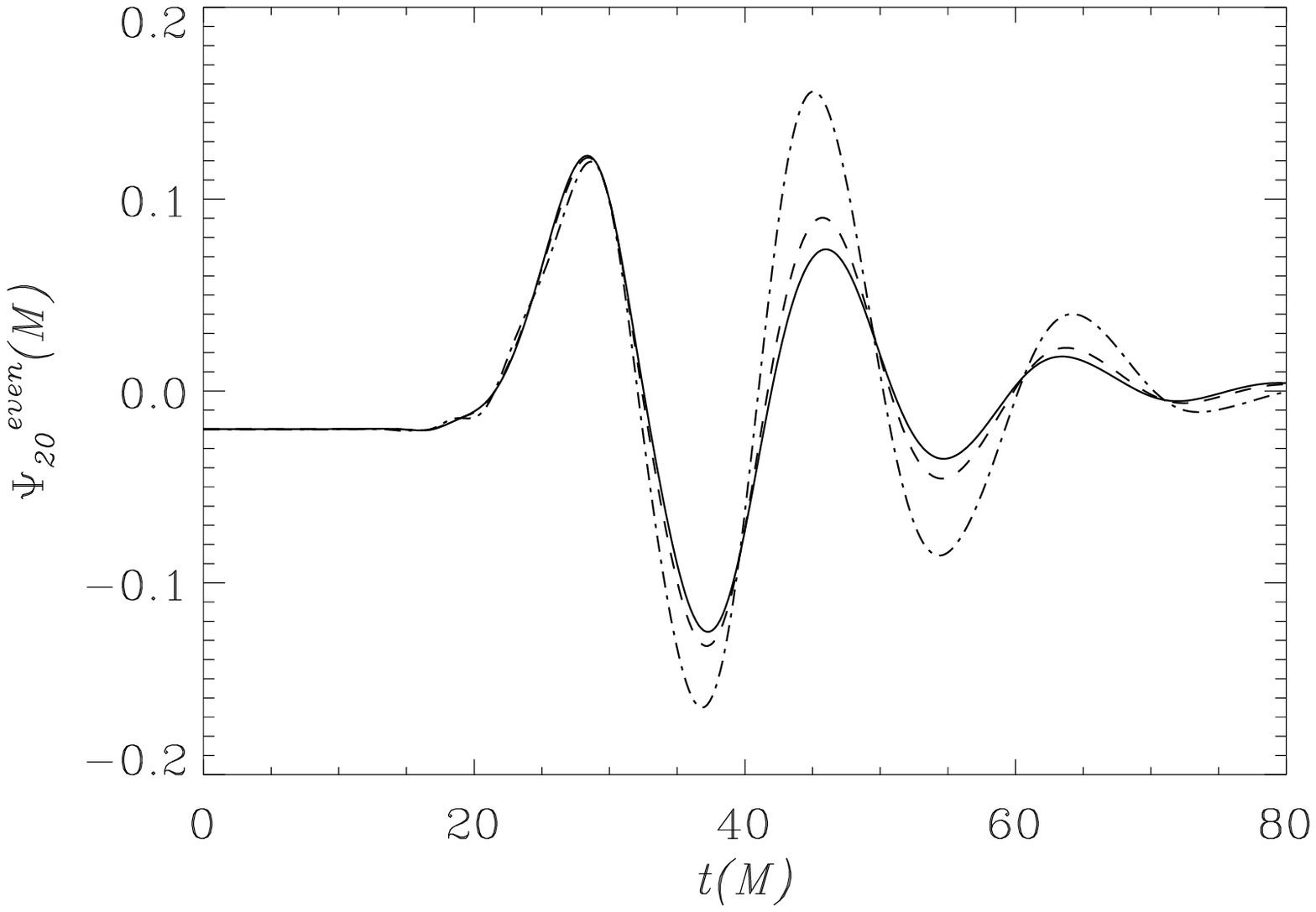}
\caption{The extracted $\ell=2$, $m=0$ waveforms for the head on collision
  of two Brill-Lindquist BH's at three different resolutions extracted at
  $18.5M$. The resolutions for the solid, dashed and dash-dotted line are 
  $0.032M$, $0.064M$ and $0.128M$ respectively.}
\label{fig:bbh0_waves}
\end{figure}

In Fig.~\ref{fig:bbh0_waves} we show the extracted $\ell=2$ and $m=0$
waveforms until $t=80 M$ for all three resolutions. The code actually
continued beyond $t=140 M$ at the highest resolution (more than $t=200
M$ at the lower resolutions) before we stopped it, due to the fact
that it was computationally fairly expensive. Initially there seem to
be some small amplitude oscillations superposed on the larger
oscillations. These seem to be related to an initial wave pulse in
the lapse moving outwards as the lapse collapses from its initial
value, which is not quite handled by the wave extraction
procedure. However these oscillations decrease with increasing
resolution. With $f = 2 \alpha^{-1} \psibl^4$ as we used in the
previous examples instead of (\ref{bbh0gauge}), the oscillations are
larger, probably because the lapse is more dynamic in the initial
phase of the evolution. But as already mentioned in
Sec.~\ref{se:lapsepuncevol}, even with $f = 2 \alpha^{-1} \psibl^0$
the lapse collapses at the punctures.  After about $t=80 M$ we see
some non-quasi-normal features in the waveform, that are most probably
due to contaminations from the outer boundary.

\begin{figure}
\epsfxsize=85mm \epsfysize=60mm \epsfbox{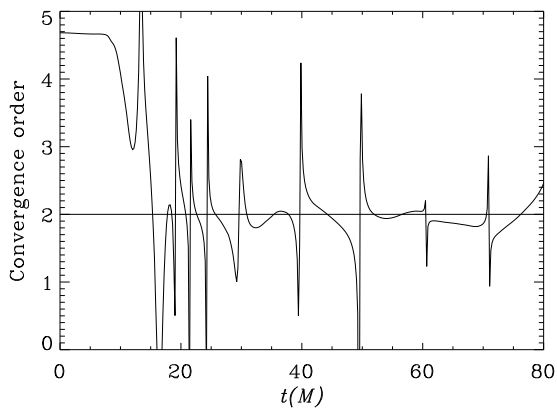} 
\caption{The convergence order in the extracted $\ell=2$, $m=0$ waveforms for
  the head on collision of two Brill-Lindquist BH's extracted at $18.5M$, based
  on the same three resolutions ($0.032M$, $0.064M$, $0.128M$) as in
  Fig.~\ref{fig:bbh0_waves}.}
\label{fig:bbh0_conv}
\end{figure}

For a wave signal $A$ extracted at three resolutions, $\Delta$, $2\Delta$
and $4\Delta$ the order of convergence, $\sigma$, can be estimated as
\beq
\sigma = \log_{2} \left| \frac{A(4\Delta)-A(2\Delta)}
                              {A(2\Delta)-A(\Delta)} \right| .
\eeq

In Fig.~\ref{fig:bbh0_conv} we show this estimate of the
convergence factor for the three waveforms from
Fig.~\ref{fig:bbh0_waves}.  Several features in this figure deserve
comment. First of all, for the first $15 M$ the signal is very small,
so the estimate of the convergence order is not very
accurate. Secondly, the phase evolution of the waveforms is somewhat
resolution dependent. This means that the curves cross over each other
at different times, leading to the spikes clearly visible in the
plot. The differences in phase evolution seem to decrease with
increasing resolution, although only at somewhere between first and
second order.  However excluding the initial part and the spikes, we
see a reasonable second order convergence in the waveforms up to
$t=80 M$.

\begin{table}
\begin{tabular}{cc}
Extremum & $\log_{2}\left |(A(4\Delta)-A(2\Delta))/
(A(2\Delta)-A(\Delta))\right |$ \\ \hline
1 & 1.17 \\
2 & 2.11 \\
3 & 2.00 \\
4 & 1.95 \\
5 & 1.96 \\
6 & 2.24 
\end{tabular}
\caption{The convergence of the amplitude for the first six local 
  extrema of the extracted $\ell=2$, $m=0$ waveforms for the head on
  collision of two Brill-Lindquist BH's extracted at radius $18.5M$.}
\label{tab:amplitude_convergence}
\end{table}

In Table~\ref{tab:amplitude_convergence}, we try to circumvent the problem
of the differently evolving phase by locating the extrema of the
waveforms and estimating the convergence order using these extremal values
even if they do not occur at the same time. As can be seen, except for the
first maximum, there is generally nice second order convergence in the 
amplitude. In the case of the first maximum, it can be seen from 
Fig.~\ref{fig:bbh0_waves} that the difference between the three 
resolutions is very small and that especially the lowest resolution is 
influenced by the pulse in the lapse moving out. 

\begin{figure}[t]
\epsfxsize=85mm \epsfysize=60mm \epsfbox{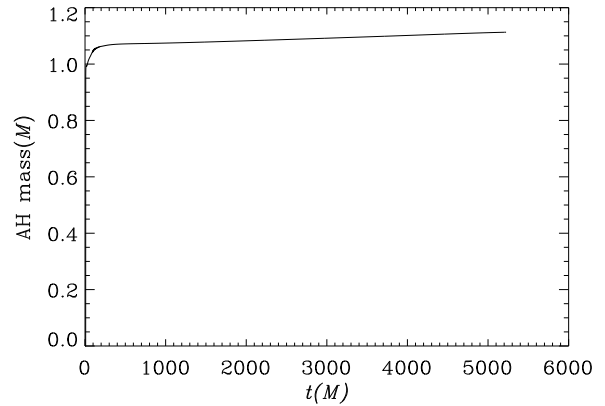} 
\caption{The apparent horizon mass for the head on collision of two 
  Brill-Lindquist BH's with maximal slicing.} 
\label{fig:bbh0_max_ah}
\end{figure}

As a second type of gauge choice we use maximal slicing and the hyperbolic
gamma driver condition with the same shift parameters as in the 1+log case,
except for the fact that $\eta = 2.0/M$ here. In this case the resolution is
$0.128M$ and the grid size is $80^3$ in octant mode. The fish-eye parameters
are $a = 4$, $s = 1.2M$ and $r_{0}=5.0M$ placing the outer boundary at a
distance of $26M$. This run ran for about a month on two processors on a
dual 1.7 GHz Xeon workstation reaching $t=5224 M$, until the machine went 
down due to an unrelated problem. 
By that time, the evolution was almost completely frozen
as can be seen from Fig.~\ref{fig:bbh0_max_ah} showing the common apparent
horizon mass as function of time. Most of the evolution occurs before
$t=200 M$ and after that there is just a slow drift of the apparent horizon
mass giving about 10\% error at $t=5000 M$.

\begin{figure}[t]
\epsfxsize=85mm \epsfysize=60mm \epsfbox{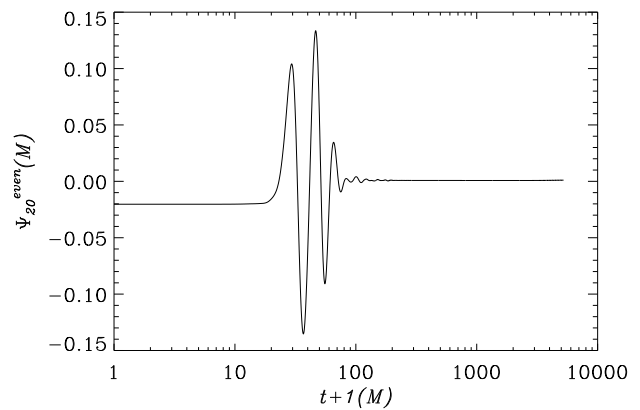} 
\caption{The extracted $\ell=2$, $m=0$ waveforms for the head on collision
  of two Brill-Lindquist BH's with maximal slicing. Note that we plot $t+1$
  in order to be able to use a logarithmic scale on the time axis.}
\label{fig:bbh0_max_wave}
\end{figure}
In Fig.~\ref{fig:bbh0_max_wave} we plot the extracted waveform with
a logarithmic time scale (actually $\ln (t+1)$) in order to be able to see
the features in the beginning of the waveform, while still showing that
it is constant and very close to zero at $t=5000 M$. The features in
the initial part of this waveform are very similar to the features in the
1+log run of the same resolution. However it is completely smooth in the
initial phase, where the 1+log waveform has the small amplitude oscillations,
since with an elliptic lapse condition, there is no wave pulse in the lapse
moving outwards.

As mentioned before, these evolutions were done in octant symmetry. We
repeated the maximal slicing evolution in bitant symmetry (reflection
about one coordinate plane), with exactly the same physical and gauge
parameters. However, this evolution died at about $t=280M$, showing
some clearly asymmetric features in the lapse and metric components in
the directions where the symmetry is not imposed. We first encountered
such a dependence of the stability of the BSSN system on the octant
symmetry in excision runs of a single black hole
\cite{Alcubierre00a}. The current result supports the conclusion that
the stability problem is not directly linked to the excision technique
or the gauge conditions, but is probably intrinsic to BSSN.  We are
currently investigating the cause of this problem.

In conclusion, with the new gauge conditions we can evolve not only
single black hole systems but also the head on collision of two black
holes with dynamically adjusting lapse and shift and reach an almost
static solution for the final black hole. While we have argued in
detail why the punctures should not evolve, and while it is plausible
that there is sufficient freedom in the gauge to almost freeze the
evolution of a single, spherical black hole, it is remarkable that the
method is successful even in the region close to and between two black
holes.


\section{Conclusion}
\label{se:conclusion}

We have discussed a new family of coordinate conditions for 3D
numerical relativity that are powerful, efficient, easy to implement,
and respond naturally to the spacetime dynamics. An application of
these conditions to previously difficult BH spacetimes shows their
strength: even without excision, they allow distorted and colliding BH
spacetimes to be evolved for more than two orders of magnitude longer
than possible previously, for thousands of $M$ rather than tens of
$M$, while keeping errors down to a few percent and allowing accurate
waveform extraction. The evolution methods and gauge choices
discussed here have already passed preliminary tests for orbiting
punctures. Work is in progress to modify the shift condition for
corotating coordinates.

\acknowledgments
We would like to thank J.~Baker for useful discussions
and comments.  The numerical experiments were implemented using BAM
and the Cactus Computational Toolkit~\cite{Bruegmann99b,Cactusweb}.
The computations were performed at the Max-Planck-Institut f\"ur
Gravitationsphysik, on the Platinum linux cluster at NCSA, on the
Hitachi SR8000-F1 at the Leibniz-Rechenzentrum of the Bavarian Academy
of Sciences and on the IBM SP at the National Energy Research Scientific
Computing Center (NERSC). This work was supported in part by 
the EU Programme `Improving the Human Research Potential and the 
Socio-Economic Knowledge Base'
(Research Training Network Contract HPRN-CT-2000-00137).


\bibliographystyle{prsty}
\bibliography{bibtex/references}

\end{document}